\documentstyle[amsmath,amsfonts,amssymb,epsfig,aps,pre,preprint]{revtex}

\title{Lorenz-like
systems and classical dynamical equations with memory forcing: 
a new point  of view for singling out the origin of chaos}
\author{R.~Festa$^{1}$, A.~Mazzino$^{2,1}$ and D.~Vincenzi$^{3,1}$}
\address{
$^1$ INFM--Department of Physics, University of Genova, I--16146
Genova, Italy\\
$^2$ CNR--ISIAtA, Polo Scientifico dell'Universit\`a, I--73100 Lecce, Italy\\
$^3$ CNRS, Observatoire de la C\^ote d'Azur, B.P. 4229,
06304 Nice Cedex 4, France}
\date{\today}            
 
\tightenlines
                     
\begin{document}

\draft
 
\maketitle

\begin{abstract}
A novel view for the emergence of chaos in
Lorenz-like systems is presented. For such purpose,
the Lorenz problem is reformulated in a classical mechanical form and
it turns out to be equivalent to the problem of a damped and forced
one dimensional motion of a particle in a 
two-well potential, with a forcing term depending on the ``memory''
of the particle past motion.
The
dynamics of the original Lorenz system in the new particle phase space
can then be rewritten in terms of an one-dimensional
first-exit-time problem.
The emergence of chaos turns out  to be due to the 
discontinuous solutions of the transcendental equation ruling the time
for the particle to cross the intermediate potential wall.
The whole problem  is tackled
analytically deriving a piecewise linearized Lorenz-like system
which preserves all the essential properties of the original model. 
\end{abstract}


\section{Introduction}
The Lorenz dynamical system,
originally introduced by Lorenz \cite{Lorenz} in order to describe in a very
simplified way the Rayleigh-B\'enard problem 
\cite{Rayleigh,Saltzman}, immediately became  important in itself as
one of the most studied low-dimensional chaotic systems. 
Still today the Lorenz model represents a paradigmatic example for both theoretical 
and numerical investigations in checking some results in chaos 
theory~\cite{Pyragas,Xu,Nijmeijer,Iplikci},
in the study of geometrical properties of dynamical 
systems~\cite{Clerc,Raab,Pingel,Yan},
in nonlinear analysis
of time series \cite{Gao,Unsworth}, 
in the stabilization and synchronization of coupled systems
\cite{Anjou,Pazo,Busch}, and so on.\\
Nevertheless, despite the great attention attracted over past decades, 
some fundamental and rigorous results have been obtained quite recently, 
as for instance the 
proof of the existence of the Lorenz attractor \cite{Tucker,Stewart},
usually using somewhat sophisticated mathematical tools.

On the contrary, our aim here is to provide a description of Lorenz system
dynamical features, which requires quite simple analytical tools and, at the
same time, allows a very intuitive inspection in Lorenz-like chaos.
Preliminary results have been reported in a short 
communication \cite{lett-EPL}. Here we shall give more details and new
results.

Our interpretation will base itself upon the fact that in the ``steady state'', i.e.,
far from the initial transient and when the memory of the initial conditions has
been lost, the Lorenz system is equivalent to a
suitably constructed second order integral-differential equation. This equation can be 
regarded, for instance, as a customary second order 
one-dimensional classical mechanics equation with
a peculiar forcing term. The corresponding dynamics can 
be interpreted as the one-dimensional motion of a particle in a 
conservative quartic two-well potential, subjected to a viscous damping 
and to an additional force resulting from the past history of the motion.
The latter force turns out to be essential for chaos to emerge as it acts
as an ``endogenous'' forcing able to permanently sustain 
the motion even in 
presence of friction.

The previous interpretation of Lorenz dynamics actually leads to a generalization
of the Lorenz model to a wider class of systems showing similar 
dynamical properties. We shall introduce 
a particular system belonging                
to such class which, because of its simplicity (piecewise linearity), will allow us 
to study 
the dynamics of the original model using analytical tools.
Indeed, in the steady chaotic regime (i.e. when the system permanently lies 
on its attractor set) the evolution of a point in the 
Lorenz phase-space can be roughly described 
as a somewhat regular amplified oscillation around one of its fixed points, 
followed by a sudden entry in the control basin of a twin fixed point
symmetrical with respect to the origin, by a new amplified oscillation around the latter
point, by a sudden return in the control domain of the former, and so on.\\
The most evident aspect of the chaotic regime  
is just the unpredictability of the instant at which the
center of the aforementioned amplified oscillations changes.
The choice of a piecewise linearized version of the original model
will allow us to highlight this point, while
keeping unchanged the peculiar topological properties
of the Lorenz dynamics. The exact equation
ruling the instant of change of the oscillation center 
will be derived and a discontinuous dependence of this instant on the initial
conditions will be highlighted.
It will also be possible to write the analytical equations that defines
the first-return two-dimensional Poincar\'e map for the piecewise linearized
system,
which in turn synthesizes the main chaotic features of the model dynamics.
Moreover, we shall show that, under suitable and reasonable conditions, the evolution of the system 
completely reduces to a one-dimensional chaotic map.\\
To summarize, starting from our interpretation,
we shall be able to propose a piecewise linearized version
of the Lorenz model, which on one hand has the same dynamical properties
of the original system and, on the other hand, will provide analytical tools
to explicate the emergence of chaos in Lorenz-like systems.

\section{The Lorenz equation}
\label{sec:1}

The original Lorenz system~\cite{Lorenz}
consists of the three first order ordinary
differential 
equations
\begin{equation}
\label{Lorenz}
\left\{  \begin{array}{lll}
\dot{X} =  -\sigma X +\sigma Y\\
\dot{Y} = -Y+(r-Z)X \\
\dot{Z} = -bZ+XY, \\
\end{array}
\right. 
\end{equation}
where the dots indicate time derivatives and $\sigma$, $b$ and $r$ are 
positive parameters
originally related to the fluid properties and to
the boundary conditions in the Rayleigh-B\'enard
problem. (Lorenz ~\cite{Lorenz} used $\sigma =10$,
$b=8/3$ and $r=28$).\\
The fixed points of system (\ref{Lorenz}) and their corresponding stabilities 
depend on $r$. For $r\leq 1$ there is only one 
(stable) fixed point in $(0,0,0)$. For $r>1$ the origin
looses its stability and a pair of new fixed
points appear: $\mbox{$(\pm[b(r-1)]^{1/2}$}, \mbox{$\pm [b(r-1)]^{1/2}$},
\mbox{$r-1)$}$, which are stable in the range $1<r\leq r_c$ with $r_c =
\sigma(\sigma+b+3)/(\sigma-b-1)$. (Note that the critical value $r_c$ exists if and
only if $\sigma>b+1$). For $r>r_c$ all fixed points are unstable, and  
the Lorenz system can 
exhibit either periodic or chaotic behavior (see, e.g., Ref. \cite{Sparrow} for
a comprehensive exposition on the matter).

Since we are interested in the case $r>1$, we can suitably define
the scaled coordinates
$x=X/[b(r-1)]^{1/2}$, $y=Y/[b(r-1)]^{1/2}$, $z=Z/(r-1)$,
so that the system  \eqref{Lorenz} becomes
\begin{equation}
\label{Lorenz scalato}
\left\{
\begin{array}{l}
\dot{x} = \sigma(y-x) \\
\dot{y} = -y+x+(r-1)(1-z)x\\
\dot{z} = b\,(xy-z)
\end{array}
\right.
\end{equation}
with fixed points $(0,0,0)$ and $(\pm 1, \pm 1 ,1)$.

We now reduce the Lorenz system to a unique differential equation for $x=x(t)$,
whose solution makes it possible the direct calculation of $y(t)$ and $z(t)$.
By inserting into the third equation of \eqref{Lorenz scalato} the
expression for $y$ in
terms of $x$ obtained from the first, one easily gets
\begin{equation}
\label{eq.z}
\dot{z}+bz=\dfrac{b}{2\sigma}\left[\dfrac{d}{dt}(x^2)+2\sigma x^2
\right],
\end{equation}
whose general solution is given by
\begin{equation}
z(t)=
e^{-b(t-t_0)}\left( z(t_0)-\dfrac{b}{2\sigma}x^2(t_0)\right)+
\dfrac{b}{2\sigma}x^2(t)+\left(1-\dfrac{b}{2\sigma}\right)
b\int_{t_0}^t ds\; e^{-b(t-s)} x^2(s).
\end{equation}
We have already stated that we are interested in the 
evolution of the Lorenz system in the chaotic steady state, i.e.
far from the initial transient. Thus, we let $t_0$ move back to $-\infty$
and obtain the steady-state expression for $z(t)$, which can be written in the form
\begin{equation}
\label{steady}
z(t)=\dfrac{b}{2\sigma}x^2(t)+
\left(1-\dfrac{b}{2\sigma}\right)
\left[x^2\right]_b(t),
\end{equation}
where we use the square brackets notation $[f]_k(t)$ to indicate the exponential
average of any
suitable time function $f$ on its past history:
\begin{equation}
[f]_k(t)\equiv
k\int_0^\infty ds\;e^{-ks}f(t-s)\;.
\end{equation}
From now on
we shall refer to $[f]_k$ as the 
\emph{$k$-exponentially vanishing memory} of the function $f$ or in short its
\emph{memory}. Note that, for the memory $[f]_k$ to exist, it suffices that
$f(t)\simeq O(\exp (-ht))$, with  $ h<k$, as $t \longrightarrow -\infty$
(the time functions we shall be dealing with are even bounded in this limit).\\
To summarize, we insert in the second equation of \eqref{Lorenz scalato}
$y(t)$ obtained from the first equation and $z(t)$ given by (\ref{steady})
and get for the variable $x=x(t)$ alone
in chaotic steady state the following 
second order differential equation with memory (in fact, an 
integral-differential equation) 
\begin{equation}
\label{eq.dinamica}
\ddot{x}+(\sigma +1)\dot{x} + \sigma (r-1)\left(\frac{b}{2\sigma}(x^2-1) x +
\left( 1- \frac{b}{2\sigma}\right)\left[x^2 - 1\right]_b x\right)\;. 
\end{equation}

One can interpret the previous equation as the dynamical equation of a (unit mass)
particle, viscously moving in a compound potential energy field consisting of a
weighted average (in the chaotic regime $b/2\sigma< 1$)  of a quartic potential energy field
independent of time (Fig. \ref{quartico})
\begin{equation}
U(x)=\sigma (r-1) \dfrac{(x^2 -1)^2}{4}
\end{equation} 
and of a quadratic potential energy field
\begin{equation}
U_t (x)=\sigma (r-1) [x^2 -1]_b \dfrac{(x^2 -1)}{2}, 
\end{equation}
whose curvature is given by a suitable memory 
function of the past motion.

Without the memory term the particle would stop in one of the minima of the bistable
potential $U$ due to the damping term $-(\sigma+1)\dot{x}$. Indeed, the
$U_t$ contribution yields, through an exponential average on the past evolution,
an ``endogenous'' forcing term which can permanently sustain the motion.
The particle oscillates with growing amplitude around the minimum of one of the potential well
until its energy is sufficient to allow the crossing of the barrier in 
$x=0$. As already said, the chaotic behavior of the system emerges just in the unpredictability 
of the instant at which the particle moves from one well to the other.
Trajectories relative to very slightly different initial conditions can produce
strongly different sequences in the number of oscillations in each well.
In the next section we shall explain this fact analytically.

In order to simplify and to make the analysis standard, 
let us rescale the time $t$ in 
Eq.~\eqref{eq.dinamica} as $t\rightarrow \tau 
\equiv[(r-1)b/2]^{1/2} \,t$, to
obtain the  equation (hereafter called \emph{Lorenz equation})
\begin{equation}
\dfrac{d^{2}x}{d\tau ^{2}}+
\eta \dfrac{dx}{d\tau }+(x^{2}-1)x=-\alpha [x^{2}-1]_
\beta\: x,
\label{eq.Lorenz}
\end{equation}
where 
\begin{displaymath}
\label{mocino}
\eta =\frac{\sigma + 1}{\sqrt{(r-1)b/2}},\qquad
\alpha=\frac{2\sigma}{b}-1,\qquad
\beta= 
\sqrt{\frac{2b}{r-1}}.
\end{displaymath}
Note that, given $b$ and $r$, 
$\sigma$ (the viscosity parameter in the original
problem) affects  both
the friction term  and the forcing term in the Lorenz equation.
This fact shows how much these ``opposite" 
contributions are in fact strictly related if the equation is to be viewed 
as representative of the
original Lorenz system. Even if one now considers the Lorenz equation as the main subject of the study,
one must note that not all the (positive) values of $\alpha$, $\beta$, $\eta$ are consistent 
with their definition in terms of the
original parameters $b$, $\sigma$, $r$. In particular,
in order to observe chaotic behavior, the following inequality must hold
\begin{equation}
\alpha> \dfrac{\eta \big( 2 + \beta (\beta + \eta )\big)}{2\beta}.
\end{equation}
(Further details on the relation between the two sets of
parameters are given in Appendix \ref{app:B}).

Equation \eqref{eq.Lorenz} allows 
us to highlight the role of the memory forcing 
term in the Lorenz system dynamics.
For this purpose
it is interesting to compare the Lorenz equation
with other examples of chaotic nonlinear (nonautonomous)
systems as, for instance, the inverse Duffing
equation $\ddot{x}+\eta\dot{x}+(x^2-1)x=A\cos (\Omega t) $,
which describes a sinusoidally forced quartic oscillator 
\cite{M-H}, or more appropriately
the parametrically forced equation
\begin{equation}
\label{unnamed} 
\ddot{x}+\eta\dot{x}+(x^2-1)x=-A\cos (\Omega t)\, x.
\end{equation}
Note that in the latter cases the motion is sustained by externally assigned forcing terms,   
while in the case of the Lorenz equation the motion is self-sustained by
the endogenous term $-\alpha \left[ x^2-1 \right]_\beta\, x$.
In Figs.~\ref{xv-Lorenz-1} 
and \ref{Duffing}
we give a numerically obtained comparison between the phase portraits of the 
Lorenz equation and of equation (\ref{unnamed}) with parameters \mbox{($\alpha$, 
$\beta$, $\eta$)},
and \mbox{($A$, $\Omega$)} suitably chosen in order to get similar ranges of motion. Note 
that the endogenous
Lorenz forcing term, mimed by $-A\cos (\Omega t)\; x$, is in fact 
neither monochromatic nor with
vanishing average value.

\section{The generalized Lorenz system}
\label{sec:2}

The Lorenz equation
\eqref{eq.Lorenz} can be usefully recast in the
form 
\begin{equation}
\label{eq.gen.}
\dfrac{d^2 x}{d\tau ^2} \,+
\, \eta \dfrac{dx}{d\tau} \,+\: 
\left(\,q(x)+\alpha\,
\left[q(x)\right]_\beta\,\right)\, \Phi'(x)\:=\,0
\end{equation}
where  
the prime indicates the derivative with respect
to $x$ and in our case
$\Phi(x)=1/2\,x^2$ and $q(x)=x^2-1$.
Such an equation can be interpreted as the description of the motion of a
unit mass particle subjected to a viscous force
$-\eta\, dx/d\tau$
and interacting with a potential field $\Phi$
through a ``dynamically varying charge''
$q_\tau (x)=q(x)+\alpha [q(x)]_\beta$. 
This charge both depends
on the instantaneous particle position $x(\tau )$ 
(by means of the term $q(x)$) and  on the past
evolution (by means of the {\em memory charge} $[q(x)]_\beta$).
The coupling of $[q]_\beta$ with the fixed potential field $\Phi$ 
acts as an endogenous forcing term which can sustain the
motion even in the presence of friction, and the chaotic 
behavior can actually arise from the synergy
between this term and the viscosity.\\
Put in the form 
\eqref{eq.gen.}, the Lorenz equation
is arranged to be generalized to a generic charge $q(x)$ 
interacting with a generic potential field $\Phi(x)$.
Correspondingly, it is possible to obtain a \emph{
generalized Lorenz system}
whose ``$x$~-~projection'' (far from the initial transient)
yields Eq.~\eqref{eq.gen.}.
Indeed, by inverting the
calculation followed to derive Eq.~\eqref{eq.gen.}
from system \eqref{Lorenz scalato}, one easily gets the generalized Lorenz dynamical system
\begin{equation}
\label{sist.gen.}
\left\{ \begin{array}{l}
\dot{x}=\sigma(y-x) \\
\\
\dot{y}=-y+x+(r-1)(1-z)\,\Phi'(x) \\
\\
\dot{z}=-bz+b[\,\frac{1}{2}
q'(x)(y-x) +q(x)+1].
\end{array}
\right.
\end{equation}
Therefore, the  specific Lorenz model can be viewed as  
singled out from a quite general class of dynamical systems
which can exhibit chaotic behavior, their common essential
property being an exponentially vanishing memory 
effect together with a viscous damping.

Equations~\eqref{eq.gen.} and \eqref{sist.gen.} are
related to  Eq.~\eqref{eq.dinamica}
by assuming as potential energy field the quantity ${\mathcal U}=(b/2\sigma)U+
(1-b/2\sigma)U_t$ 
with
\begin{equation} 
\begin{array}{lcr}
U=\sigma (r-1)\int q(x)\,\Phi'(x)\, dx 
&\qquad\text{and}\qquad& 
U_t=\sigma (r-1)[q]_\beta \,\Phi\; .
\end{array} 
\end{equation}
Obviously, any choice of $q$ and $\Phi$ 
should  maintain
the main properties of the Lorenz model,
i.e., correspond to
a two-well piecewise differentiable potential energy $U(x)$, such that $U(x)\to\infty$ as
$|x|\to\infty$.

We shall now 
focus our attention on a particular choice for $q$ and $\Phi$, that 
will maintain all the qualitative properties of the Lorenz
system and, at the same time, will allow us to deal with 
chaos analytically.

\section{The piecewise linearized Lorenz system}
\label{sec:3}

\subsection{Linearization near fixed points}

As already noticed, the chaotic behavior of the
Lorenz system essentially depends on the unpredictability
of the instants when $x$ change its sign: as long as it
keeps constant sign the system evolution is 
certainly nonlinear, and nevertheless not ``chaotic" at all.
This fact suggests a slight modification of the original
form of the Lorenz system, in order to single out analytically the origin
of chaos without to be faced with the difficulties
arising from nonlinear features. We thus set in
Eqs.~\eqref{eq.gen.} and \eqref{sist.gen.} $\Phi (x)=|x|$ and 
$q(x)=|x|-1$ obtaining (for $x\neq 0$) the piecewise linear Lorenz-like equation
\begin{gather}
\label{eq.lin.}
\frac{d^2 x}{d\tau^2}+\eta \frac{dx}{d\tau}+ 
\left\{|x|-1 
+\alpha \left[\,|x|-1\,\right]_{\beta} \,\right\}
\mbox{sgn}(x)=0,\\
\intertext{where $\text{sgn}(x)\equiv |x|/x$. 
The corresponding piecewise linearized
dynamical system, with the original choice of parameters, is then given by }
\label{sist.lin.}
\left\{ \begin{array}{l}
\dot{x}=\sigma (y-x) \\
\\
\dot{y}=-y+x +(r-1)(1-z) \mbox{ sgn}(x)\\
\\
\dot{z}=-bz+b\,\mbox{sgn}(x)\dfrac{x+y}{2}.
\end{array}
\right.                                     
\end{gather}
Our assumptions on $\Phi$ and $q$ correspond in Eq.~\eqref{eq.dinamica} 
to $U(x)=\sigma (r-1)
\left(\,|x| - 1 \,\right)^2 /2$   and
$U_t(x)= \sigma (r-1) 
\left[\,|x| - 1\,\right]_b
 \left(\, |x|-1\,\right)$. Thus, we are faced with a simplified model, obtained
by replacing the constant-in-time quartic potential with a piecewise quadratic
one resulting from the superposition of 
two parabolas with vertex in $\pm 1$ and truncated at $x=0$
(Fig.~\ref{parabole}). The
two-well character of $U$ is obviously maintained as well as 
the piecewise differentiability.
The replacement of the original potential actually corresponds
to a linearization of the system around both unstable fixed points
$(\pm 1, \pm1,1)$, with the metching performed in $x=0$.
It appears that the chaotic behavior of the original model
does not depend on the differentiability in  $x=0$.
One can guess that also other classes of dynamical systems
can be transformed in a piecewise linearized version by means
of the same operations.

It is easy to check that
the fixed points of \eqref{sist.lin.} are $(\pm 1, \pm 1, 1)$ and that
the equation ruling their local stability is the same as for the original
Lorenz system with parameters $b$, $\sigma$, $\rho\equiv(r+1)/2$. 
In particular, if $\sigma>b+1$, 
the critical value of $r$
for our piecewise linear system \eqref{sist.lin.} is given by
 $r_c^{(lin)}=2r_c-1$. 
In Figs.~\ref{xv-Lorenz-2} and \ref{xv-Lorenz-Lin} two chaotic phase portrait 
for systems~(\ref{Lorenz scalato})
and (\ref{sist.lin.}) are shown, corresponding to the same
choice of $b$ and $\sigma$, and different choices of $r$ in order to preserve
the relationships between $r$ and the proper corresponding critical value.\\   
When dealing with 
system \eqref{sist.lin.},
the main simplification is that it is
separately linear for 
$x<0$ and $x>0$ and  can thus be  analytically
solved in each region. 
Indeed, by applying the operator $(d/d\tau + \beta)$ to each side
of Eq. \eqref{eq.lin.} one obtains the equation 
\begin{equation}
\label{lin.x}
\dfrac{d^3 x}{d\tau^3}+(\beta+\eta)\dfrac{d^2 x}{d\tau^2}+
(1+\beta\eta)\dfrac{dx}{d\tau}+\beta(1+\alpha)(x-
\text{sgn}(x))=0 \;,
\end{equation}
which can be explicitly solved separately on each side of $x=0$. 
The nonlinearity of the model is  simply reduced to
a change of sign of the forcing term $\pm \beta (1+\alpha)$
when $x$ crosses the plane $x=0$, henceforth denoted
with $\pi$ \cite{nota:feedback}. 
As we shall see, the crossing times are somewhat unpredictable, as they
result from the discontinuous solutions of an (incidentally transcendental) equation.
Our piecewise linearized system will thus turn out to be an important
tool to analytically investigate the emergence of chaos in
Lorenz-like systems. The main advantage of the piecewise linearization
is that one has to deal only with the simplest
nonlinearity, i.e.
an isolated (not eliminable)
discontinuity (see, e.g., Refs.~\cite{Andronov,Sparrow1,Holmes,Chua,Chua2} 
for other examples of piecewise linear chaotic dynamical
systems).

\subsection{Analysis of the motion}

Let us now consider in detail the
second order 
integral-differential equation~\eqref{eq.lin.},
which describes the evolution in time of $x$ for the piecewise
linearized system in the chaotic steady state.
It is equivalent to a third order nonlinear differential equation 
whose phase space is described by the coordinates
$x$, $\dot{x}$, $\ddot{x}\,$ 
(with a small abuse of notation from this time on
we shall indicate with the dot
the differentiation with respect to $\tau$).
Solutions of
Eq.~\eqref{eq.lin.} can be  
easily calculated  with tools of the customary analysis on each side
of $x=0$. To obtain a global solution, 
such partial solutions should be matched at $x=0$ 
under the reasonable assumptions  
that the position $x$, 
the velocity $\dot{x}$,
and the memory $[\,|x|\,]_\beta$ are
continuous
\cite{nota:contin}. Notice that,
in contrast, when crossing the plane $\pi$,
the acceleration $\ddot{x}$ turns out to be undefined.  
However, if $x(\tilde{\tau})=0$, referring to the left and right time limits 
$\ddot{x}(\tilde{\tau}^-)$ and $\ddot{x}(\tilde{\tau}^+$), 
it appears from Eq. \eqref{eq.lin.} that they are related by the equation
$\ddot{x}(\tilde{\tau}^-) 
+\ddot{x}(\tilde{\tau}^+)
=-2\eta\,\dot{x}(\tilde{\tau}). 
$
In the sequel we shall refer
to $\ddot{x}(\tilde{\tau}^-)$ 
and $\ddot{x}(\tilde{\tau}^+)$ as the acceleration ``immediately'' before
and ``immediately'' after the crossing, respectively.

Since
Eq.~\eqref{eq.lin.} is invariant under the transformation
$x \rightarrow -x$, $\tau\rightarrow\tau$, we can focus our attention only 
on one
of the two regions,
e.g. $x>0$, and describe the motion
in this half-space (the evolution in its twin half-space being recovered through the change
$x \rightarrow -x$). Clearly, this is equivalent to put a ``rigid wall'' 
in $x=0$ and look at the crossing of $\pi$ as
an elastic collision. \\
In summary,
according to the previous scheme, the system evolution in time is 
completely described by the following steps:
\begin{enumerate}
\item motion for $x>0$;
\item collision against $\pi$ and discontinuity
of $\ddot{x}$;  
\item inversion $x\rightarrow -x$ and matching with a 
new solution defined in the region $x>0$ again.
\end{enumerate}

\subsubsection{Motion in the half-space $x>0$}

If we define $\xi\equiv x-1$, Eq.~\eqref{lin.x} assume a simple form
\begin{equation}
\label{eq.3}
\dddot{\xi}+(\beta+\eta)\,\ddot{\xi}\,+\,(1+\beta\eta)\,\dot{\xi}
\,+\,\beta(1+\alpha)\,\xi=0\: ,
\end{equation}
that is a linear third order differential
equation (homogeneous and with constant coefficients).
As a consequences of the 
Routh-Hurwitz theorem \cite{Lancaster},
a critical value $\alpha_c$ for the parameter $\alpha$ exists, i.e.
\begin{equation}
\alpha_c = \frac{(1+\beta \eta)(\beta + \eta)}
{\beta}-1. 
\end{equation}
For $\alpha>\alpha_c$ the fixed points are 
unstable saddle focus with a real negative eigenvalue
$-\lambda_0$ and a pair of complex conjugates eigenvalues
$\lambda_1=\lambda_r+i\lambda_i$, $\bar{\lambda}_1=\lambda_r-i\lambda_i$. 

For the sake of simplicity we shall indicate
with $(0,\tau_1)$ the time interval between two 
consecutive collisions in the steady-state of the system.
Without loss of generality
we assume $x=0$ and $\dot{x}>0$. Then, it can be easily shown  
that in the time interval 
$(0,\tau_1)$ the motion in the phase space
$\xi$, $\dot{\xi}$, $\ddot{\xi}$  ($\xi>-1$)
is completely described by
the equation 
\begin{equation}
\label{eq.matr.}
\boldsymbol{\xi}(\tau)\, =\, 
      {\cal M}(\tau)\,{\cal M}(0)^{-1}\, 
     \boldsymbol{\xi}_0\, ,
\end{equation}
where we have defined 
\begin{gather*}
      \boldsymbol{\xi}(\tau)\equiv
      \begin{pmatrix}
            \xi (\tau)\\
            \dot{\xi}(\tau)\\ 
            \ddot{\xi}(\tau)
      \end{pmatrix},
\qquad\qquad\quad       
      \boldsymbol{\xi}_0\equiv
      \begin{pmatrix}
            \xi_0\\
            \dot{\xi}_0\\ 
            \ddot{\xi}_0
      \end{pmatrix}
\equiv
\begin{pmatrix}
      -1\\
      \dot{\xi}(0)\\[0.1cm] 
      \ddot{\xi}(0^+)
\end{pmatrix},
\\
\intertext{and}
{\cal M}(\tau)\:=\:
\begin{pmatrix}
\mbox{Re} \left(e^{\lambda_1 \tau}\right) & \mbox{Im}  
\left(e^{\lambda_1 \tau}\right) &  e^{-\lambda_0 \tau}\\  \\
\mbox{Re}\left(\lambda_1 \, e^{\lambda_1 \tau}\right) & 
\mbox{Im}  
\left(\lambda_1\,e^{\lambda_1 \tau}\right) &  
-\lambda_0 \,e^{-\lambda_0 \tau}\\  \\
\mbox{Re} \left(\lambda_1^2 \, e^{\lambda_1 \tau}\right) & 
\mbox{Im} 
\left(\lambda_1^2\,e^{\lambda_1 \tau}\right) &  
\lambda_0^2 \,e^{-\lambda_0 \tau}
\end{pmatrix}\, . 
\end{gather*} 
(Note that, 
for $\alpha>\alpha_c$,
${\mathcal M}(0)$ is always invertible
since $\text{det}\,{\cal{M}}(0)=       
\lambda_i \left[ \lambda_i ^2 +
(\lambda_r +\lambda_0)^2 \right]$ and 
$\lambda_i >0$).
The eigenvalues of the matrix ${\mathcal M}(\tau)\,
{\mathcal M}(0)^{-1}$, which connects the vector
$\boldsymbol{\xi}(\tau)$ to its initial value $\boldsymbol{\xi}_0\,$, 
are  
$e^{\lambda_1 \tau}$,
$e^{\bar{\lambda}_1 \tau}$, 
$e^{-\lambda_0 \tau}$, with corresponding (constant-in-time) 
eigenvectors 
\begin{displaymath}
\begin{array}{lcr}
{\boldsymbol v}_1\equiv
  \begin{pmatrix}
  1\\ \lambda_1 \\[0.05cm] \lambda_1 ^2
  \end{pmatrix},&
\boldsymbol{v}_2\equiv
  \begin{pmatrix}
  1\\ \bar{\lambda}_1 \\[0.05cm] \bar{\lambda}_1 ^2
  \end{pmatrix},&
\boldsymbol{v}_3\equiv
  \begin{pmatrix}
  \ 1\\ -\lambda_0 \\[0.05cm]\quad \lambda_0^2
  \end{pmatrix}
  \end{array} .
\end{displaymath}  
With respect to the base
$\left\{
\boldsymbol{v}_1,
\boldsymbol{v}_2,\boldsymbol{v}_3\right\}$ 
one has
\begin{equation}
\label{eq.diag.}
\boldsymbol{\xi}(\tau)=
e^{\lambda_r\tau}\left(
c_1\,
e^{\lambda_i\tau}\boldsymbol{v}_1+c_2\,
e^{-\lambda_i\tau}\boldsymbol{v}_2\right)+c_3\,
e^{-\lambda_0\tau}\boldsymbol{v}_3\,.
\end{equation}
For the saddle point $\boldsymbol{\xi}=
\boldsymbol{0}$     
there exist a stable one-dimensional manifold 
${\mathcal W}^s$
corresponding to  $\boldsymbol{v}_3$
\begin{gather*}
{\mathcal W}^s=
\big\{
p\,\boldsymbol{v}_3\ | \ p>-1 \big\}\\
\intertext{
and an unstable two-dimensional manifold
${\mathcal W}^u$  generated by 
$\boldsymbol{v}_1$,
$\boldsymbol{v}_2$ (see Fig.~\ref{sella})}
{\mathcal W}^u=
\left\{
p\, \dfrac{\boldsymbol{v}_1+\boldsymbol{v}_2}{2}\,+\,
q\,\dfrac{\boldsymbol{v}_1-\boldsymbol{v}_2}{2i}
\quad | \quad p>-1\ ,\; q\in \mathbb{R}
\right\}.  
\end{gather*}
The evolution of
the linearized system
in the interval $(0,\tau_1)$  is the combination
of an exponential decay along ${\mathcal W}^s$ and 
of an amplified rotation on ${\mathcal W}^u$.
To determine the relative quickness of each component
of the motion with respect to the other, consider  that
from Eq.~\eqref{eq.3} it is easily checked that
$\lambda_0-2\lambda_r=\beta+\eta$, and so
$\lambda_0>\lambda_r$ 
$\forall \ \alpha,\,\beta,\,\eta$.
Therefore, the exponential decay along the stable
manifold ${\mathcal W}^s$ is always more rapid than the exponential oscillating growth
on the unstable manifold ${\mathcal W}^u$, the parameter controlling
this difference being $\beta +\eta$.\\
Although the fixed points are saddle focus, it appears that the piecewise
linearized system does not exhibit Shilnikov chaos~\cite{Shilnikov}
owing to the absence of an homoclinic orbit.

\subsubsection{$\pi$-collision and inversion}

To complete the description of the piecewise linearized system dynamics,
we consider the instant $\tau_1$ at which 
the first collision occurs.
At this time, as already remarked, the trajectory coming from
the half-space $x>0$ must be matched with 
the solution defined again in the same region,
but with corresponding new ``initial conditions'' \cite{nota:condiz.}
\begin{subequations}
\begin{gather}
\label{condiz.iniz.}
\left\{
  \begin{array}{lllll}
  \xi_1&=&-1\\ 
  \dot{\xi}_1&=& -\dot{\xi}(\tau_1)\\
  \ddot{\xi}_1&=&-\ddot{\xi}(\tau_1^+)&=&
  \ddot{\xi}(\tau_1^-)+2\eta\,\dot{\xi}(\tau_1),
  \end{array}
  \right.   \\
\intertext{or, in matrix form,} 
\label{urto}
\boldsymbol{\xi}_1\,=\,
{\mathfrak I}\,{\mathcal D}\,
\boldsymbol{\xi}(\tau_1^+) 
\end{gather}
\end{subequations}
with 
\begin{eqnarray*}
{\mathcal D}\equiv
\begin{pmatrix}
1&0&0\\
0&1&0\\
0&-2\eta&-1
\end{pmatrix}  &\qquad \text{and} \qquad&
{\mathfrak I}\equiv
\begin{pmatrix}
1&0&0\\
0&-1&0\\
0&0&-1                        
\end{pmatrix}.                         
\end{eqnarray*}
The matrix
${\mathcal D}$ accounts for the acceleration discontinuity
in $x=0$, while ${\mathfrak I}$ yields
the sign inversion after the impact. 
From 
Eqs.~\eqref{eq.matr.} and \eqref{urto} it follows that
the velocity and the acceleration immediately
after the  collision are related to the initial conditions
by the operator ${\mathcal P}(\tau)\equiv{\mathfrak I}
\,{\mathcal D}\,{\mathcal M}(\tau)\,{\mathcal M}(0)^{-1}$,
according to the formula
\begin{equation}
\label{P}
\boldsymbol{\xi}_1={\mathcal P}\big(\tau_1(\boldsymbol{\xi}_0)\big)\,
\boldsymbol{\xi}_0\, .
\end{equation}
Notice 
the highlighted dependence of $\tau_1$ on $\boldsymbol{\xi}_0$, 
which reveals
the nonlinear character of this important relationship. 
  
Starting from the above results, the very origin of chaos
in the piecewise linearized Lorenz system will be identified
and discussed in the next section. We shall also show that the 
basic mechanisms for chaos to emerge apply also to the 
original Lorenz system.

\subsection{Dependence of $\tau_1$ on initial conditions}

\subsubsection{Unpredictability of the crossing time} 

As already observed, the instant at which the plane $\pi$
is crossed, strongly depends on the initial conditions.
This fact is strictly related to the chaotic behavior
of Lorenz-like systems. Let us now study in some detail this
topic for the piecewise linearized model.

The instant $\tau_1$ at which the first collision occurs
is defined by the condition $\xi (\tau_1)=-1$. 
From the first line of Eq.~\eqref{eq.matr.} $\tau_1$ is thus 
the smallest positive
solution of the transcendental equation
\begin{equation}
\label{trasc.}
-1
=C_1\, e^{\lambda_r \tau_1}\cos (\lambda_i \tau_1)\, +\,
   C_2 \,e^{\lambda_r \tau_1}\sin (\lambda_i \tau_1)\, +\,
   C_3\, e^{-\lambda_0 \tau_1} 
\end{equation}
where $C_1$, $C_2$, $C_3$ are linearly related to initial conditions. 
The {\em residence time} $\tau_1$ is therefore the first intersection of the graphs 
of $g(\tau_1)= -C_3\, e^{-\lambda_0 \tau_1} -1$ 
and $h(\tau_1)= 
C_1\, e^{\lambda_r \tau_1}\cos (\lambda_i \tau_1)\, +\,
   C_2\, e^{\lambda_r \tau_1}\sin (\lambda_i \tau_1)$.
Since $g$ is a decreasing exponential function and
$h$ an oscillating function with growing amplitude,
one can easily understand why  even a little 
modification of initial conditions can produce 
a discontinuous variation of $\tau_1$ (Fig.~\ref{intersez}). 

In conclusion, 
the unpredictability of the
residence time is  closely connected to the discontinuous 
character of  the solutions of Eq.~\eqref{trasc.}. The chaotic behavior for the 
the piecewise linearized  system clearly stems from such unpredictability.
Our claim is that an entirely analogous situation exists also for the original Lorenz model.
We shall return however in more detail on this important analogy in the sequel.

\subsubsection{The residence time  $\tau_1$ 
as a  function of $\dot{\xi}_0$ and 
$\ddot{\xi}_0$}
\label{sec:Graph}

We now explicitly investigate the dependence of $\tau_1$ on
the initial velocity  and  acceleration.
As previously remarked, the function
$\tau_1=\tau_1 (\dot{\xi}_0,\,
\ddot{\xi}_0 )$ is implicitly defined
as the smallest positive solution of Eq.~\eqref{trasc.}.
Since
$C_1$, $C_2$, $C_3$ are linearly related
to the initial conditions, Eq.~(\ref{trasc.}) can be rewritten
in terms of $\dot{\xi}_0$ and
$ \ddot{\xi}_0$. In this form it  describes  a
family ${\mathcal S}$ of  straight lines in the plane $\{\dot{\xi} _0,\ddot{\xi}_0\}$ 
parametrized by $\tau_1$
\begin{equation}
\label{fascio S}
{\mathcal S}:\qquad A(\tau_1)\,\dot{\xi} _0\: + \:
B(\tau_1)\,\ddot{\xi}_0\: + \: C(\tau_1)\,=\,0
\end{equation}
[see Appendix \ref{app:fascio} for the explicit expression
of the coefficients $A(\tau_1)$,  $B(\tau_1)$,  $C(\tau_1)$].\\
Let us denote with
$(T _i ), \;{i\geq 0}$ the ordered sequence
of the zeros of $B(\tau_1)$.
Note that one always has $T_0 =0$. 
The slope of the straight lines of $\mathcal{S}$,
$-A/B$, and their ordinate for $\dot{\xi}_0 =0$, i.e. $-C/B$,
have their singularities in   $T_i,\,i\geq 0$.
Both these functions are only  
asintotically periodic because of the presence of terms
proportional to  $e^{-\lambda_0 \tau_1}$, which
become negligible only for large $\tau_1$ 
(Figs.~\ref{coeff} and \ref{intercetta}).\\
In each interval 
$(T_i,T_{i+1})$, $i\geq 0$, the function 
$-A/B$ is everywhere growing (see Appendix~\ref{app:fascio})
and varies from $-\infty$
and $+\infty$. Therefore, as $\tau_1$ increases, the straight
lines of $\mathcal{S}$ rotate anticlockwise and at the same 
time translate starting from the $\ddot{\xi_0}$-axis 
(obtained for $\tau_1=0$)
(Fig.~\ref{ragnatela}).\\
The envelope of the family $\mathcal{S}$ is a
curve
$$\gamma : \begin{cases}
         \dot{\xi}_0=\dot{\xi}_0(\tau_1)\\
         \ddot{\xi}_0=\ddot{\xi}_0(\tau_1)  
         \end{cases}$$
which  looks similar to  an elliptic spiral
(Fig.~\ref{inviluppo}). (A parametrical
representation of $\gamma$ is given in Appendix~\ref{app:fascio}).

From Eq.~\eqref{fascio S} the contour lines
of the function $\tau_1$ are parts of straight lines in the
plane $(\dot{\xi}_0, \ddot{\xi}_0 )$. Indeed, 
according to the definition, each line
of $\mathcal{S}$  is in fact a contour line for $\tau_1$ only 
where its points do not belong to another line
corresponding to a smaller value of $\tau_1$ and, furthermore,
only where their abscissas correspond to positive values of $\dot{\xi}_0$.
Rather than a family of straight lines, the
function contour lines are raies
or segments according to the constant value of $\tau_1$.\\
For $\tau_1$ ranging from $0$ to $T_2$ these lines
perform a complete anticlockwise ``rotation,''
starting from the \mbox{$\ddot{\xi}_0$-axis}. Thus, they  cover the entire
half-plane $\dot{\xi}_0\geq 0$ except  the region
inside their envelope; from this first rotation
one obtains a set of raies (Fig.~\ref{semirette}).\\
Subsequently, for $\tau_1$ ranging from $T_2$ to $T_4$ 
the contour lines become segments
of variable orientation and contained in the region 
delimited by the curve already generated from the first 
rotation and the new envelope of the second set
of straight lines. 
This behavior repeats itself $\forall\: \tau_1 \in 
[T_{2i},T_{2i+1}]$, $i\geq 0$.
Because of the aperiodic character of the functions 
involved, the curve $\gamma$ outlines in the positive
half-plane a structure consisting of ``pseudo-elliptic''
annula.
Moving anticlockwise along each of these annula,
$\tau_1$ grows continuously. On the contrary, passing through 
the border that separates two different bands, one
meets  discontinuities in the dependence of
$\tau_1$ on initial conditions.\\
The curve $\gamma$ ``winds'' round the point $P_0
\equiv(\lambda_0,-\lambda_0^2)$ (see Fig.~\ref{inviluppo}). 
For $(\dot{\xi}_0,
\ddot{\xi}_0)\to (\lambda_0,-\lambda_0^2)$
one has $\tau_1\to\infty$, since for these initial
conditions one obtains $C_1=C_2=0$ and 
the system exactly lies on the stable manifold
${\mathcal W}^s$. Its motion is in this case an exponential 
decay towards the fixed point $\boldsymbol{\xi}=
\boldsymbol{0}$. 

The previous observations allow to easily guess the structure
of the graph of $\tau_1=\tau_1(\dot{\xi}_0,\ddot{\xi}_0)$
shown in Fig.~\ref{piramide}.
It should be noted that $\tau_1$ shows instability
with respect to the initial conditions only in a 
limited subset of the half-plane $\dot{\xi}_0\geq 0$.
As we shall see,
it is natural to expect  that in the chaotic regime the system
is quickly attracted inside this region.

\subsection{The piecewise linearized model 
as a one-dimensional map}

\subsubsection{The $\pi$-plane Poincar\'e map}

At this point of our study we have analyzed in some detail the 
unpredictability of the time at which the system crosses the plane $\pi$.
To completely motivate the chaotic dynamics of the piecewise linearized system
and thus of the original model, we must, however, add some further results 
on the attracting set of the system.

All  results  obtained in the previous section
can be easily extended to the \mbox{$n$-th} collision against $\pi$.
Specifically, denoting with $\boldsymbol{\xi}_n$ 
the array assigning position, velocity,
and acceleration immediately after the \mbox{$n$-th}
collision
$$\boldsymbol{\xi}_n\equiv
\begin{pmatrix}-1\\ \dot{\xi}_n\\
\ddot{\xi}_n\end{pmatrix},$$
we have [similarly to Eq.~\eqref{P}]
\begin{eqnarray}   \label{mappa}
\boldsymbol{\xi}_n&=&
{\mathcal P}(\tau_n)\, \boldsymbol{\xi}_{n-1},
\end{eqnarray}
where
we have denoted with $\tau_n$ the \mbox{$n$-th} residence time, i.e.,
the time interval between the
\mbox{($n-1$)-th} collision and the \mbox{$n$-th} one.\\
The first line
of Eq.~\eqref{mappa} gives $\tau_n$ in terms of
$\dot{\xi}_{n-1}$ and 
$\ddot{\xi}_{n-1}$: the dependence
of this time on the \mbox{$(n-1)$-th} initial conditions has already been
discussed in Section~\ref{sec:Graph} .
Knowing $\tau_n$,  the other two lines 
allow to relate
the velocity $\dot{\xi}_n$
and the acceleration 
$\ddot{\xi}_n$ to 
$\dot{\xi}_{n-1}$ and $\ddot{\xi}_{n-1}\,$.
Eq.~\eqref{mappa} defines a two-dimensional first-return
Poincar\'e map between $(\dot{\xi}_n,
\ddot{\xi}_n)$
and $(\dot{\xi}_{n-1},\ddot{\xi}_{n-1})$,
obtained from the section $\xi=-1$ of
the phase space $\xi,$ $\dot{\xi}$, $\ddot{\xi}$.
We recall that, in spite of its  appearance, the map~\eqref{mappa} is in fact non-linear, 
since the matrix
$\mathcal P$ depends on $\tau_n$, which is a
transcendental discontinuous function of the $(n-1)$-th
initial conditions.\\
Indeed, under suitable conditions, the two-dimensional map for the linearized system
practically reduces to a one-dimensional map.

\subsubsection{Approximation of the system with a one-dimensional map}

It has been
already shown that for $\xi >-1$ the motion 
in the phase space has the form~\eqref{eq.diag.}
$$
\boldsymbol{\xi}(\tau)=c_1\,
e^{\lambda_1\tau}\boldsymbol{v_1}+c_2\,
e^{\bar{\lambda}_1\tau}\boldsymbol{v_2}+c_3\,
e^{-\lambda_0\tau}\boldsymbol{v_3}$$
where $\lambda_0$ and $\lambda_r=\text{Re}(\lambda_1)$ 
satisfy the equation
$\lambda_0-2\lambda_r=\beta+\eta$, which in turn implies
$\lambda_0>\lambda_r$. 
As a consequence the evolution in the phase space
$\xi$, $\dot{\xi}$, $\ddot{\xi}$ consists both
of a ``rapid'' decay towards $\boldsymbol{\xi}=
\boldsymbol{0}$ along the stable manifold 
${\mathcal W}^s$ and of a ``slow'' amplified
oscillation on ${\mathcal W}^u$. 
We thus expect the phase trajectories
to be strongly attracted on the unstable manifold
and, once on ${\mathcal W}^u$, to slowly spiral outwards.
Provided trajectories start close enough to
${\mathcal W}^u$, then they meet the $\pi$-plane  
very close to its intersection with ${\mathcal W}^u$
itself, i.e., along the straight line 
\begin{equation}
{\mathcal L^-}:
\begin{cases}
\ddot{\xi}+2\lambda_r\dot{\xi}+(\lambda_r^2+
\lambda_i^2)=0\\
\xi=-1.
\end{cases}
\end{equation}
Let us assume that the attraction towards the unstable manifold
is very strong 
and thus it takes place almost instantaneously
(the goodness of this assumption
is controlled by $\beta+\eta$). Under this hypothesis all
trajectories approximately hit $\pi$ along the straight line $\mathcal{L}^-$.
Thus, from Eq.~\eqref{urto}, it follows that, immediately
after each collision, the system necessary lies very close to the
straight line  
\begin{equation}
{\mathcal L^+}: 
\begin{cases}
\ddot{\xi}+2(\eta+\lambda_r)\dot{\xi}-(\lambda_r^2+
\lambda_i^2)=0\\
\xi=-1. 
\end{cases} 
\end{equation}
Therefore, the following relation between velocity 
and acceleration is expected to approximately hold
\begin{equation}
\ddot{\xi}_n+2(\eta+\lambda_r)\,\dot{\xi}_n 
-(\lambda_r^2+
\lambda_i^2)= 0\, .
\label{a-v}
\end{equation} 
From Eqs. \eqref{eq.lin.} and \eqref{a-v}  an analogous
linear dependence between the velocity and the memory follows
\begin{equation}
\label{vel-mem}
(\eta+2\lambda_r)\dot{\xi}_n-\alpha\,
w_n+1
-(\lambda_r^2+
\lambda_i^2)=0\; ,
\end{equation}
where $w_n$ denotes
the \mbox{$\beta$-memory} of $\xi$ evaluated at the \mbox{$n$-th} collision. 
As a consequence,
in the limit we have considered, 
the attracting set for the map \eqref{mappa} is ${\mathcal L}^+$ 
and, therefore, it reduces to an one-dimensional map.

Since in the chaotic steady-state the velocity and the
acceleration that define the initial conditions for
the trajectory after each impact are not independent,
the time $\tau_n$ can be expressed as a function of 
$\dot{\xi}_{n-1}$ alone: $\tau_n=
\tau_n(\dot{\xi}_{n-1})$ (see Fig.~\ref{t-v}).
The behavior of this map is easily understood if we refer to the
graph of $\tau_1$ as a functions of $\dot{\xi}_0$ and 
$\ddot{\xi}_0$ (see Fig.~\ref{piramide}). Indeed,
in the chaotic regime $\mathcal{L}^+$ is superimposed to
the region 
of the plane $(\dot{\xi}_{n-1},
\ddot{\xi}_{n-1})$
contained by $\gamma$, where $\tau_1$ shows unstable behavior
with respect to a change in the initial conditions (Fig.~\ref{sovrapp}).
For those values of $\dot{\xi}_{n-1}$,
for which $\mathcal{L}^+$ passes through the same ``pseudo-elliptic''
corona, $\tau_n$ slowly changes with varying crossing
velocity. On the contrary, when $\mathcal{L}^+$  intersects the
boundary between two different coronas,
$\tau_n$ shows a discontinuity in its dependence on 
$\dot{\xi}_{n-1}$ (Fig.~\ref{t-v}). \\
The linear dependence between velocity
and acceleration immediately after each collision
implies that 
the Poincar\'e map \eqref{mappa}
becomes one-dimensional, e.g. a map between 
$\dot{\xi}_n$ and $\dot{\xi}_{n-1}$.
Figure~\ref{v-v} shows this fact for
$\alpha =6.50$, $\beta=0.19$, $\eta=0.78$.
By simple inspection of the derivatives corresponding to the fixed points
of the map
one can  easily check that they are all unstable, so that the map
produces a chaotic behavior. Note that the discontinuities simply correspond
to the analogous ones
in the function  $\tau_n=
\tau_n(\dot{\xi}_{n-1})$, evaluated along the straight line
given by \eqref{a-v} and drawn in Fig.\ref{sovrapp}.

For the sake of completeness,
we turn now to the description of 
the piecewise linearized system
in the original phase space $x$, $\dot{x}$, $\ddot{x}$. There the system
has two fixed points $\left(
\pm 1,0,0\right)$.
For each one there exists a stable manifold
\begin{gather}
{\mathcal W}^s_{\pm}=
\big\{
p\,\boldsymbol{v_3}+\,
(\pm 1,0,0)
\quad | \quad p \gtrless \mp 1 
\big\}\\
\intertext{and an unstable manifold}
{\mathcal W}^u_\pm=
\left\{
p\,\dfrac{\boldsymbol{v_1}+\boldsymbol{v_2}}{2}\,+\,
q\,\dfrac{\boldsymbol{v_1}-\boldsymbol{v_2}}{2i}\,+\,
(\pm 1,0,0)
\quad | \quad p \gtrless \mp 1\ 
\text{and}\ q\in {\mathbb R}
\right\} \; .  
\end{gather}
During the evolution the trajectories are rapidly
attracted on the unstable manifold relative to the half-plane
where they belong (Fig.~\ref{spaziofasi}). 
Thus, immediately after the crossing of the
$\pi$-plane the phase point lies on one of the straight lines
\begin{gather}
\begin{cases}
\ddot{x}+2(\eta+\lambda_r)\dot{x}+
(\lambda_r^2+\lambda_i^2)\,\text{sgn}(\dot{x})=0\\
x=0,\end{cases}
\nonumber \\
\intertext{while the memory and the velocity 
are linearly related according to equation}
(\eta+2\lambda_r)\,\dot{x}_n\,-\,\alpha\,
\text{sgn}(\dot{x}_n)\,
w_n\,+\,1
-(\lambda_r^2+
\lambda_i^2)=0\; .
\label{mem}
\end{gather}
As for the crossing of the $\pi$-plane, the system can therefore 
be described by a one-dimensional map.
To get the maps $\tau_n=\tau_n(\dot{x}_{n-1})$ and
$\dot{x}_n=\dot{x}_n(\dot{x}_{n-1})$
from the analogous ones in the rigid
wall scheme, one has simply to consider that $\dot{\xi}_n$ 
coincides with $|\dot{x}_n|$. 

The reader should note that all the maps we have drawn in the 
previous figures have been obtained by numerically solving the
analytical equation~\eqref{mappa}, which exactly defines $\tau_n$,
$\dot{\xi}_{n+1}$, and $\ddot{\xi}_{n+1}$, and 
they have not been computed by numerical integration of the differential system 
(as usual for non solvable dynamical systems like the Lorenz original one).
Thus, the use of numerical tools has been required only because of the transcendental
character of the concerned equation.

We explicitly remark that in general the attraction towards ${\mathcal W}^u$ is not infinite,
but nevertheless more rapid than the amplified rotation on 
the unstable manifold (remember that $\lambda_0-2 \lambda_r=
\beta+\eta>0$ $\forall \ \alpha,\beta,\eta$).
Therefore, we can expect that, if $\beta+\eta$ is finite,
the attractor set is not exactly the straight line ${\mathcal L}^+$,
but a narrow ``strip'' that contains  ${\mathcal L}^+$.
This strip intersects the region of the graph of $\tau_n$ where
the residence time is strongly dependent on initial conditions and the mechanism 
for chaos to arise is absolutely the same.
Our numerical simulations for finite $\beta+\eta$ 
are in excellent agreement with these predictions
and then we do not show their corresponding graphs here, since they are practically
indistinguishable from the graphs we have already discussed. 

\section{Comparison with the original Lorenz system}
\label{sec:4}

We conclude our analysis by showing some simulated numerical
results. Bfore doing that, some remarks are worth discussing.
We previously considered $x,\dot{x},\ddot{x}$
as independent coordinates in the phase space because of
their physical meaning: 
it is more intuitive to speak about acceleration
rather than of memory of the system. Unfortunately, if we want to
``assign initial conditions'' to the original Lorenz system after the 
crossing of the plane $\pi$, we cannot consider
the acceleration and the velocity since from Eq.~\eqref{eq.Lorenz}
they are not independent variables in $x=0$.
Indeed the relation $\ddot{x}_n+\eta\,\dot{x}_n=0$
always holds, while the memory is not known \emph{a priori}.
As a consequence, we have to consider as independent initial conditions
position, velocity, and memory.
However, it is worth remarking that
the comparison between the 
Lorenz system and the piecewise linearized one
is completely meaningful.
This
 because for the latter model
$w_{n-1}$ is easily known in terms of $\dot{\xi}_{n-1}$ 
and $\ddot{\xi}_{n-1}$ from Eq.~\eqref{eq.lin.}:
for the linearized system the choice of 
 $\ddot{\xi}_{n-1}$ or of $w_{n-1}$ is absolutely equivalent.

Since we are interested in the behavior of the Lorenz system 
in the chaotic steady state, all the following results are thus obtained 
via the numerical integration of the steady-state
differential system
\begin{equation}
\label{steady-state}
\begin{cases}
\dot{x}=v\\
\dot{v}+
\eta\, v+(x^2-1)x=-\alpha \,w\, x\\
\dot{w}=-\beta \, w +\beta (x^2-1),
\end{cases}
\end{equation}
where now $w = [x^2 - 1]_\beta$. 
This dynamical system, as already proved, is equivalent to the system~\eqref{Lorenz scalato}
far from the initial transient.

We can now turn to the comparison with the piecewise linearized system.
For the original Lorenz model we have computed the maps that relate $\tau_n$
and $|\dot{x}_n|$ to $|\dot{x}_{n-1}|$, respectively,
by numerical integration of the system \eqref{steady-state}
(the notations are the same as in the previous section).
Figures~\ref{t-vL} and \ref{v-vL} show the results we obtained:
the qualitative behavior of these maps is very similar to that of the 
analogous maps for
the piecewise linearized system. As we will see, the mechanism for chaos
to arise is indeed the same.\\
Observe that the maps in Figs.~\ref{t-vL} and \ref{v-vL}
are not perfectly onto: obviously this is due to the fact that, in general, the
Lorenz system does not reduce exactly to an one-dimensional map when
the plane $x=0$ is crossed.

From direct computation it appears that the residence time around a fixed point 
depends on the initial conditions in a way similar
to that of the linearized system (Fig.~\ref{torreL}).
There is a region in which $\tau_1$ varies continuously with varying
initial conditions and a region (Figs.~\ref{torreL} and
\ref{torreL1}) of strong instability.
In the latter region $\tau_1$ changes continuously along each
``annulus'' of the ``spiral,'' whereas it shows abrupt discontinuities
crossing the boundary of each annulus. As expected, the system
is attracted on that region (Fig.~\ref{memL}) and the
maps in Figs.~\ref{t-vL} and \ref{v-vL} derive from the
overplotting between Figs.~\ref{memL} and \ref{torreL1}.\\
The chaos actually emerges due to the combination
of the step-like first-exit-time (Fig.~\ref{t-vL}) and of the 
piecewise return map (Fig.~\ref{v-vL}) for initial 
conditions. This mechanism is the same as for the linearized
model, for which we gave an analytical description.
The results we obtained for the piecewise linearized system
therefore provide in our opinion a useful enlightening tool to understand
both the original Lorenz system and its eventual generalizations.

Finally, in order to point out the connection between the original Lorenz
system and its piecewise linearized version, it is illuminating
to compare the tent-like map shown by Lorenz in his original work~\cite{Lorenz}
with the analogous map for our simplified system. Describing the relation
between two consecutive maxima of the coordinate $z(t)$, Lorenz found
a univocal tent-like map (Fig.~\ref{tenda}).
For our linearized system we obtain similar results (see Fig.~\ref{tenda-lin}). 
The growing branch of the map is approximately linear, since for
small values of $z$ the system evolution is confined in a half-space
and, there, it essentially consists
of an exponentially amplified oscillation. For large values of $z$
the map has the same shape of the Lorenz tent-like map.
The linear growing of the left branch of the map
is due to the simplification we made on the evolution
on each side of $x=0$. The shape of the right branch
shows that the piecewise linear system keeps all the complexity 
of the Lorenz model, which can be attributed to the unpredictability
of the $x=0$ crossing,
which is the very origin of chaos in Lorenz-like systems.

\section{Conclusions}

The main conclusions from the present paper can be summarized as follows.
We have reformulated in a classical mechanics form the
dynamics of Lorenz-like systems and showed that
its three-dimensional phase-space dynamics can be 
mapped into an one-dimensional
motion of a particle oscillating in a conservative quartic
two-well potential, subjected to a viscous dissipation
and to a memory forcing.
Starting from this interpretation, we have introduced a
piecewise linearized version of the Lorenz system 
(belonging to a larger family
of Lorenz-like systems) which
substantially has the same properties of the original model with 
the advantage that it allows an analytical treatment.
\\
The most evident aspect of chaotic regime is the unpredictability
of the instant at which the center of the particle amplified 
oscillation changes. Chaos arises due to the combination
of the step-like behavior of this time with the piecewise
return map defining the crossing conditions.
This aspect has been singled out analytically, focusing on the
piecewise linearized version. There, the exact equation for the
time, at which the oscillation center changes, has been derived
and the discontinuous dependence of this time on the crossing 
conditions has been shown analytically.\\
By means of numerical simulations we have verified that the
highlighted mechanisms for the chaos to emerge survive also for
the fully-nonlinear Lorenz system, where analytical techniques are
not applicable.

\acknowledgments

We would like to thank M. La Camera and A. Vulpiani for useful
discussions and suggestions.
This work was partially supported by the INFM project GEPAIGG01.
D. V. was partially supported
by the doctoral grants of the University
of Nice and by grants of the University of Genova.

\appendix

\section{Parameters $\alpha$, 
$\beta$, $\eta$}
\label{app:B}
In the Lorenz system, the following conditions for $\sigma,\,b,\,r$
have to be satisfied in order 
to have  three instable fixed points:
\begin{subequations}
\begin{eqnarray}
& &\sigma,\,b,\,r \in \mathbb{R}^+,\label{prima} \\
& &r>1,  \label{seconda}\\
& &\sigma>b+1,\label{terza}  \\
& &r>r_c\; .  \label{quarta}
\end{eqnarray}
\end{subequations}
The parameters $\alpha,\,\beta,\,\eta$ are defined as follows:
$$\alpha=\frac{2\sigma}{b}-1,\qquad 
 \beta=\sqrt{\frac{2b}{r-1}}\,,\qquad 
\eta =\frac{\sigma + 1}{\sqrt{(r-1)b/2}}. 
$$
From (\ref{prima}) and (\ref{seconda})
it immediately follows that
$\alpha,\,\beta,\,\eta$ are positive too. 
Therefore, $\sigma, \, b,\, r$ can be rewritten in terms of
$\alpha,\,\beta,\,\eta$ in the form 
\begin{gather*}
\nonumber
\sigma=-\frac{(1+\alpha)\beta}{\beta(\alpha+1)- 2\eta},\\[0.2cm]
r=1-\frac{4}{\beta(\beta(\alpha+1)- 2\eta)},\\[0.2cm]
\nonumber
b=-\frac{2\beta}{\beta(\alpha+1)- 2\eta}.
\end{gather*}
Under the hypothesis $\alpha,\,\beta,\,\eta \in \mathbb{R}^+$,
(\ref{prima}) and (\ref{seconda}) 
are equivalent to 
\begin{gather}
\label{diseq1}
\beta(\alpha+1)<2\eta, \\
\intertext{while (\ref{terza}) corresponds to the
inequality}
\label{dise2}
 \beta(\alpha +1 )>\beta + \eta \,  .\\
\intertext{Specifically, (\ref{prima}),
(\ref{seconda}), and (\ref{terza}) then imply}
\label{ultima}
\alpha>\frac{\eta}{\beta}>1\, .\\
\intertext{Finally, (\ref{quarta}) can be rewritten
in terms
of new parameters as}
\alpha>\dfrac{\eta \big(2+\beta(\beta + \eta)\big)}{2\beta}
\,  ,
\end{gather}
from which necessarily \eqref{ultima} follows.\\

\section{Family $\mathcal{S}$} 
\label{app:fascio}

The family of straight lines $\mathcal{S}$ 
is defined by  the equation
\begin{gather*}
\label{fascio}
{\mathcal S}:\qquad A(\tau_1)\,\dot{\xi} _0\: + \:
B(\tau_1)\,\ddot{\xi}_0\: + \: C(\tau_1)\,=\,0
\end{gather*}
where
\begin{displaymath}
\begin{array}{lll}
A(\tau_1)
&=&
-2\,\lambda_i\,\lambda_r\,  e^{-\lambda_0 \tau_1}\,+\,
e^{\lambda_r \tau_1} \left(
                      2\,\lambda_i\,\lambda_r\,
                       \cos (\lambda_i \tau_1)\,+
                      \,(\lambda_0^2 +\lambda_i^2 -\lambda^2_r)\,
                      \sin(\lambda_i \tau_1)\,
                 \right)\\
\\
B(\tau_1)&=&                 
\lambda_i\, e^{-\lambda _0 \tau_1} \,+\, 
     e^{\lambda _r  \tau_1}
      \left(\, \left( \lambda _r +
           \lambda _0 \right) 
        \sin (\lambda _i \tau_1)\,
          -\lambda_i\, \cos (\lambda _i \tau_1)\, 
              \right)\\
\\
C(\tau_1)&=&
-\lambda_i\,(\lambda_i^2 +\lambda_r^2)\, e^{-\lambda_0 \tau_1}\,-\,
\lambda_0\,e^{\lambda_r \tau_1}\big[\:
\lambda_i\,(2\,\lambda_r+\lambda_0)\,\cos(\lambda_i \tau_1)\,+\\
\\
\ & \ &
+\,(\lambda_i^2-(\lambda_0+\lambda_r)\lambda_r)\,\sin(\lambda_i \tau_1)
\:\big]
+\, 
(\,\lambda_i\,(\lambda_i^2 +(\lambda_0 + \lambda_r)^2\,). 
\end{array}        
\end{displaymath} 
 
\subsection{Slope of straight lines of $\mathcal{S}$}

We indicate with
$(T _i )_{i\geq 0}$ the ordered sequence
of the zeros of $B\,$: $        
T_i <T_{i+1} \ \linebreak\forall \ i\geq 0$
with
$B(T_i)=0   \ \forall \ i\geq 0 .
$
The first derivative of $-A/B$ is positive over 
the whole domain of definition. Indeed one has
\begin{gather*}
-\dfrac{d}{d\tau_1}\left( \dfrac{A}{B} \right)= \\ \\
\dfrac{\lambda_i e^{\lambda_r \tau_1}\big[
\lambda_i e^{\lambda_r \tau_1}+ e^{-\lambda_0 \tau_1}\big(
-\lambda_i \cos (\lambda_i \tau_1) -(\lambda_0+\lambda_r )
\sin (\lambda_i \tau_1) \big) \big] 
\left(\lambda_i^2 +(\lambda_0+\lambda_r)^2\right)}
{ 
\lambda_i e^{-\lambda_0 \tau_1}+ e^{\lambda_r \tau_1}\big[
-\lambda_i \cos (\lambda_i \tau_1) - (\lambda_r+\lambda_0 )
\sin (\lambda_i \tau_1) \big]^2  
}
\end{gather*}
with
\begin{displaymath}
\lambda_i e^{\lambda_r \tau_1}+ e^{-\lambda_0 \tau_1}\big(
-\lambda_i \cos (\lambda_i \tau_1) - (\lambda_0+\lambda_r )
\sin (\lambda_i \tau_1)\big)  \, >\,0 \qquad
\forall \; \tau_1\in 
\displaystyle{\bigcup_{i=0}^{\infty}}(T_i,T_{i+1})
\subset\mathbb{R}^{+}.
\end{displaymath} 
The latter inequality can be easily proved. Indeed:
\begin{gather*}
(\lambda_r+\lambda_0)\tau_1 > \log (1 + (\lambda_r +\lambda_0)\tau_1)
\;\Rightarrow \\
e^{(\lambda_r+\lambda_0)\tau_1} > 1+(\lambda_r +\lambda_0)\tau_1 >
\cos (\lambda_i \tau_1) + \dfrac{(\lambda_r +\lambda_0)}{\lambda_i}\tau_1
\;\Rightarrow \\
\lambda_i e^{\lambda_r \tau_1} > e^{-\lambda_0 \tau_1} \big(  \lambda_i
\cos(\lambda_i \tau_1) + (\lambda_r +\lambda_0)\sin (\lambda_i \tau_1) 
\big )\qquad
\forall \; \tau_1\in 
\displaystyle{\bigcup_{i=0}^{\infty}}(T_i,T_{i+1})
\subset\mathbb{R}^{+}. 
\end{gather*}

\subsection{Envelope of family $\mathcal{S}$} 

A parametrical (not regular) representation
of the envelope of the family $\mathcal{S}$ 
is obtained 
deriving $\dot{\xi}_0$ and 
$\ddot{\xi}_0$ with respect to $\tau_1$. This can be done  from the 
system
$$ \left\{ \begin{array}{l}
A(\tau_1)\,\dot{\xi} _0\: + \:
B(\tau_1)\,\ddot{\xi}_0\: + \: C(\tau_1)\,=\,0 \\[0.2cm]
\partial_{\tau_1} A(\tau_1)\:\dot{\xi} _0\; + \;
\partial_{\tau_1}B(\tau_1)\:\ddot{\xi}_0\; + \; \partial_{\tau_1}C(\tau_1)\,=\,0  
\; .   
\end{array}\right. $$
The result 
$$\gamma : \begin{cases}
         \dot{\xi}_0=\dot{\xi}_0(\tau_1)\\
         \ddot{\xi}_0=\ddot{\xi}_0(\tau_1)  
         \end{cases} 
$$
is given by
\begin{multline*}
\dot{\xi}_0(\tau_1)=\\[0.15cm] \big\{
-\lambda_0\big( \lambda_i(1+e^{-\lambda_0 \tau_1})
\cos(\lambda_i \tau_1)-\lambda_i(e^{-(\lambda_0+\lambda_r)\tau_1}+
e^{\lambda_r \tau_1})
\big)\\[0.15cm]
-(\lambda_r^2 + \lambda_i^2-\lambda_0\lambda_r) 
(e^{-\lambda_0 \tau_1}-1)\sin(\lambda_i \tau_1) 
\big\}
\bigg/ \\[0.15cm]
\big( \lambda_i e^{\lambda_r \tau_1} + e^{-\lambda_0 \tau_1}
(-(\lambda_0 +\lambda_r)\sin(\lambda_i \tau_1)-\lambda_i
\cos(\lambda_i \tau_1)\big)
\end{multline*} 
and
\begin{multline*}
\ddot{\xi}_0(\tau_1)=\\[0.15cm] 
\big\{ \lambda_0^2 \big(
\lambda_i(\cos(\lambda_i \tau_1)-e^{\lambda_r \tau_1})
     +\lambda_r\sin(\lambda_i \tau_1)\big)\\[0.15cm] 
+(e^{-\lambda_0 \tau_1}-1)(\lambda_i\cos(\lambda_i \tau_1) 
-\lambda_r \sin(\lambda_i \tau_1))(\lambda_r^2 + \lambda_i^2)+
\\[0.15cm] 
-\lambda_0\big(
e^{-\lambda_0 \tau_1}((\lambda_i^2-\lambda_r^2) \sin(\lambda_i \tau)+
2\lambda_i\lambda_r\cos(\lambda_i \tau_1))
-2 \lambda_i\lambda_r e^{-(\lambda_0+\lambda_r) \tau_1} 
\big)\big\}\bigg/
\\[0.15cm] 
\big( \lambda_i e^{\lambda_r \tau_1} + e^{-\lambda_0 \tau_1}
(-(\lambda_0 +\lambda_r)\sin(\lambda_i \tau_1)-\lambda_i
\cos(\lambda_i \tau_1)\big).
\end{multline*}


\begin{figure}[!h]
\caption{The constant-in-time
quartic potential $U$ ($\sigma=10$, $b=8/3$, $r=28$).
The classical particle representing the Lorenz system
moves in the potential $U$ subjected to a viscous damping and 
to a memory forcing. The minima of the potential wells
correspond to the unstable points of the three-dimensional
Lorenz system.}
\label{quartico} 
\end{figure}

\begin{figure}[!h]
\caption {Typical chaotic phase portrait
for the original Lorenz model 
($\eta=1.31$, $\alpha=10.30$,
$\beta=0.216$).}
\label{xv-Lorenz-1} 
\end{figure}

\begin{figure}[!h]
\caption{Typical chaotic phase portrait for the inverse
parametrically forced Duffing equation
($\eta=1.31$, $A=4.2$, $\Omega=0.99$).
The chaotic evolution of this system is characterized
by the competition between a viscous friction and a forcing term
with the same structure of the right hand side of the Lorenz equation. Here,
however, the forcing is externally given and the memory
is replaced by a known function of time.}
\label{Duffing} 
\end{figure}

\begin{figure}[!h]
\caption {Quartic potential $U$ for the original
Lorenz system (dashed line) and for the piecewise linearized system
(full line). The linearization of the Lorenz system maintains
the qualitative shape of the constant-in-time potential.}
\label{parabole} 
\end{figure}

\begin{figure}[!h]
\caption{Typical chaotic phase portrait for the original
Lorenz system
($\sigma =10$, $b=8/3$, $r=28$).}
\label{xv-Lorenz-2} 
\end{figure}

\begin{figure}[!h]
\caption{Typical chaotic phase portrait for the
piecewise linearized Lorenz system
($\sigma=10$, $b=8/3$, $r=55$).}
\label{xv-Lorenz-Lin} 
\end{figure}

\begin{figure}[!h]
\caption{Stable and unstable manifolds for the
point $\boldsymbol{\xi}=\boldsymbol{0}$ in the phases space
$\xi$, $\dot{\xi}$, $\ddot{\xi}$.
The evolution of the piecewise linearized system on each side
of $\pi$ consists of an exponential decay along ${\mathcal W}^{s}$
and of an amplified oscillations on ${\mathcal W}^{u}$.}
\label{sella}
\end{figure}

\begin{figure}[!h]
\caption{Graphical interpretation of the discontinuous character of 
$\tau_1$ for small changes of the initial condition. Full line 
is the curve $g(\tau_1)$, dashed lines is $h(\tau_1)$
for the parameters $\alpha =6.50$, $\beta=0.19$, $\eta=0.78$.
Bullets denotes the first intersection between $g$ and $h$, whose abscissa 
defines the residence time.}
\label{intersez}
\end{figure}

\begin{figure}[!h]
\caption{Slope 
of the straight lines of the family $\mathcal{S}$
($\alpha =6.50$, $\beta=0.19$, $\eta=0.78$).
[See Eq.~\eqref{fascio S} and Appendix \ref{app:fascio} for the 
exact definition.]}
\label{coeff}
\end{figure}

\begin{figure}[!h]
\caption{Ordinate for $\dot{\xi}_0 =0$
of the straight lines of the family $\mathcal{S}$
($\alpha =6.50$, $\beta=0.19$, $\eta=0.78$).
[See Eq.~\eqref{fascio S} and Appendix \ref{app:fascio} for the 
exact definition.]} 
\label{intercetta} 
\end{figure}

\begin{figure}[!h]
\caption{The family of curves $\mathcal{S}$ 
[defined in Eq.~\eqref{fascio S}] for 
$\alpha =6.50$, $\beta=0.19$, and $\eta=0.78$.}
\label{ragnatela} 
\end{figure}
 
\begin{figure}[!h]
\caption{The envelope $\gamma$ of the family of straight lines 
$\mathcal S$ looks similar to a spiral 
($\alpha =6.50$, $\beta=0.19$, and $\eta=0.78$).}
\label{inviluppo}
\end{figure}

\begin{figure}[!h]
\caption{Contour lines of the residence time  $\tau_1$
as a function of the initial conditions ($\dot{\xi}_0$,
$\ddot{\xi}_0$) for 
$\tau_1 \in [0,T_2]$
($\alpha =6.50$, $\beta=0.19$, $\eta=0.78$).}
\label{semirette} 
\end{figure}

\begin{figure}[h]
\caption{Graph of $\tau_1$ as a function of the initial conditions,
obtained by numerical solution of the transcendental equation~\eqref{trasc.}
($\alpha=6.50$, $\beta=0.19$, $\eta=0.78$).}
\label{piramide}
\end{figure}

\begin{figure}[!h]
\caption{Map of the first-exit time $\tau_n$ as a function
of the crossing velocity $\dot{\xi}_n$, obtained by numerical
solutions of Eqs.~\eqref{mappa} and \eqref{a-v}.
($\alpha =6.50$, $\beta=0.19$, $\eta=0.78$)}
\label{t-v}
\end{figure}

\begin{figure}[h]
\caption{During chaotic evolution the couple of crossing 
conditions $(\dot{\xi}_n,\ddot{\xi}_n)$ is attracted
on the straight line ${\mathcal L}^+$, which
is superimposed to the instability region of the
residence time $\tau_n$ (identified by the curve $\gamma$).
The graph refers to the values
$\alpha =6.50$, $\beta=0.19$, $\eta=0.78$.}
\label{sovrapp}
\end{figure} 

\begin{figure}[!h]
\caption{The velocity $\dot{\xi}_n$ at the \mbox{$n$-th} crossing
of the plane $x=0$ as a function of the velocity $\dot{\xi}_{n-1}$
($\alpha =6.50$, $\beta=0.19$, $\eta=0.78$).
Notice that, however, in the steady-state chaotic evolution 
the map is restricted to the range $(0,1.2)$
of possible values of $\dot{\xi}_{n-1}$.}
\label{v-v}
\end{figure}

\begin{figure}[!h]
\caption{Attraction of the piecewise linearized 
system on the manifolds ${\mathcal W}^u_\pm$.
The evolution of the system on each side of $\pi$ consists of
a ``rapid'' exponential decay along the stable manifold
and of a  ``slow '' amplified oscillation on the unstable manifold.
}
\label{spaziofasi}
\end{figure} 

\begin{figure}[!h]
\caption{Map $\tau_n=
\tau_n(|\dot{x}_{n-1}|)$ for the original 
Lorenz system, obtained by
numerical integration of the system~\eqref{steady-state} 
($\alpha =10.30$, $\beta=0.216$, $\eta=1.31$).}
\label{t-vL}
\end{figure}

\begin{figure}[!h]
\caption{Map between $|\dot{x}_n|$ and
$|\dot{x}_{n-1}|$
for the original Lorenz system, obtained by
numerical integration of the system~\eqref{steady-state} 
($\alpha =10.30$, $\beta=0.216$, $\eta=1.31$).}
\label{v-vL}
\end{figure}

\begin{figure}[!h]
\caption{Graph of 
$\tau_1$ as  a function of the crossing conditions
for the original Lorenz system, obtained by
numerical integration of the system~\eqref{steady-state} 
for different initial conditions
($\alpha =10.30$, $\beta=0.216$, $\eta=1.31$),
}
\label{torreL}
\end{figure}

\begin{figure}[!h]
\caption{Graph of $\tau_1$ as a function of the crossing conditions
viewed from the above in the case of the original Lorenz system
($\alpha =10.30$, $\beta=0.216$, $\eta=1.31$).}
\label{torreL1}
\end{figure}

\begin{figure}[!h]
\caption{Map between the crossing memory $w_n$ and
the corresponding crossing velocity $|\dot{x}_n|$
for the original Lorenz system
($\alpha =10.30$, $\beta=0.216$, $\eta=1.31$).}
\label{memL}
\end{figure}

\begin{figure}[!h]
\caption{The tent-like map for the original Lorenz system
($b=8/3$, $\sigma=10$, $r=28$). $M_n$ denotes the 
$n$-th maximum of the coordinate $z(t)$.}
\label{tenda}
\end{figure}
 
\begin{figure}[!h]
\caption{The tent-like map for the piecewise linearized system
($b=8/3$, $\sigma=10$, $r=100$).}
\label{tenda-lin}
\end{figure}


\newpage
\begin{center} 
\psfig{file=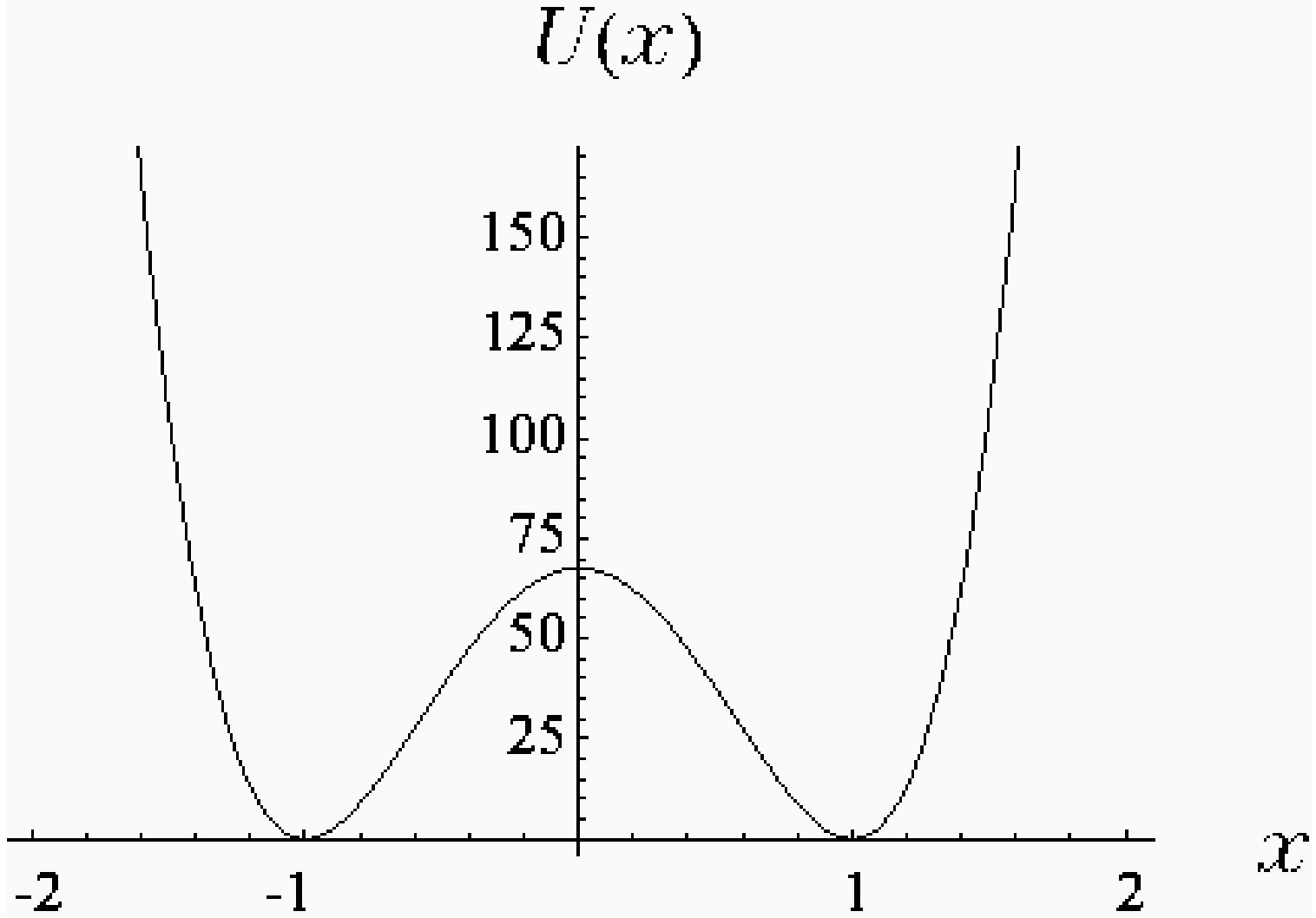,width=8.6cm} 
\\
FIG. \ref{quartico}. R. Festa \emph{et al.}, Phys. Rev. E
\end{center}

\newpage
\begin{center} 
\psfig{file=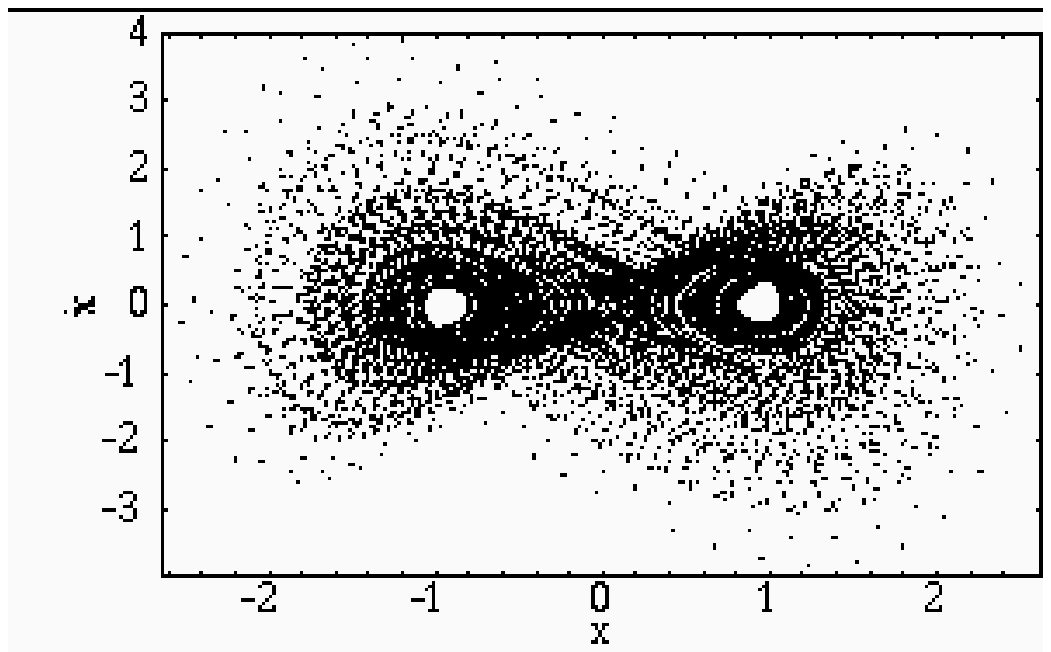,width=8.6cm} 
\\
FIG. \ref{xv-Lorenz-1}. R. Festa \emph{et al.}, Phys. Rev. E
\end{center}

\newpage
\begin{center} 
\psfig{file=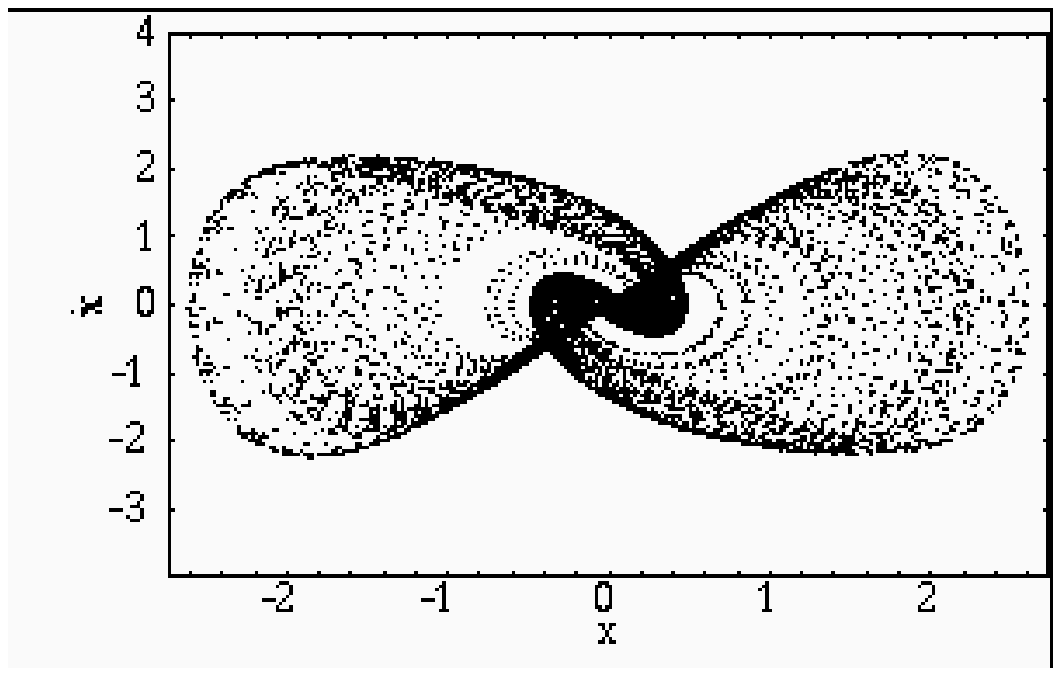,width=8.6cm} 
\\
FIG. \ref{Duffing}. R. Festa \emph{et al.}, Phys. Rev. E
\end{center}

\newpage
\begin{center} 
\psfig{file=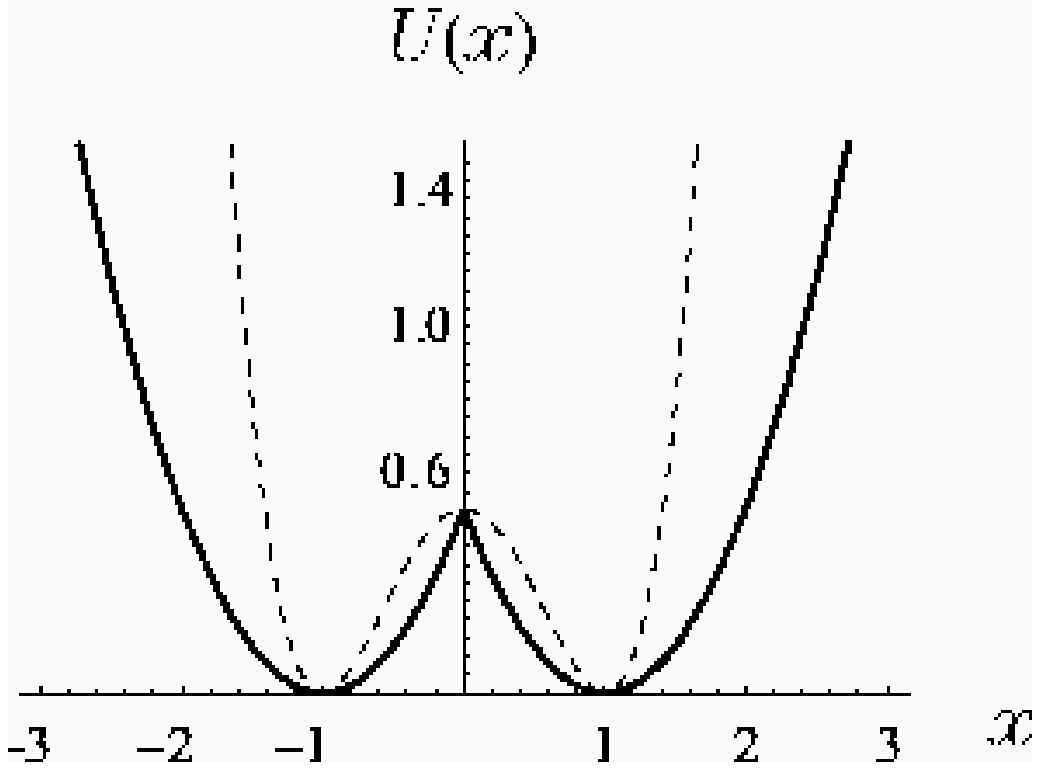,width=8.6cm} 
\\
FIG. \ref{parabole}. R. Festa \emph{et al.}, Phys. Rev. E
\end{center}

\newpage
\begin{center} 
\psfig{file=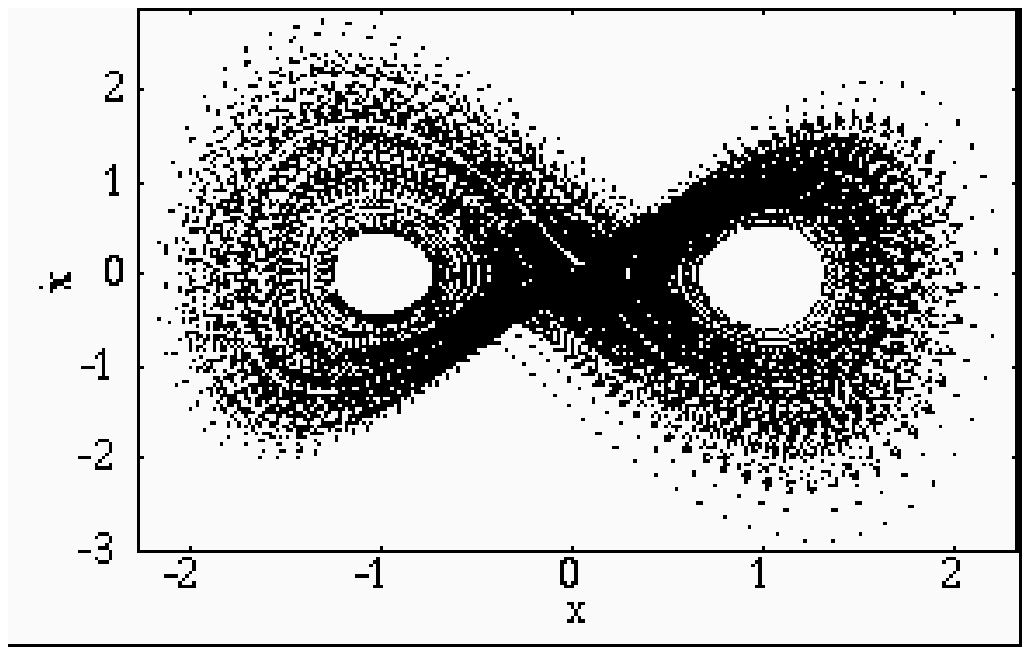,width=8.6cm} 
\\
FIG. \ref{xv-Lorenz-2}. R. Festa \emph{et al.}, Phys. Rev. E
\end{center}

\newpage
\begin{center} 
\psfig{file=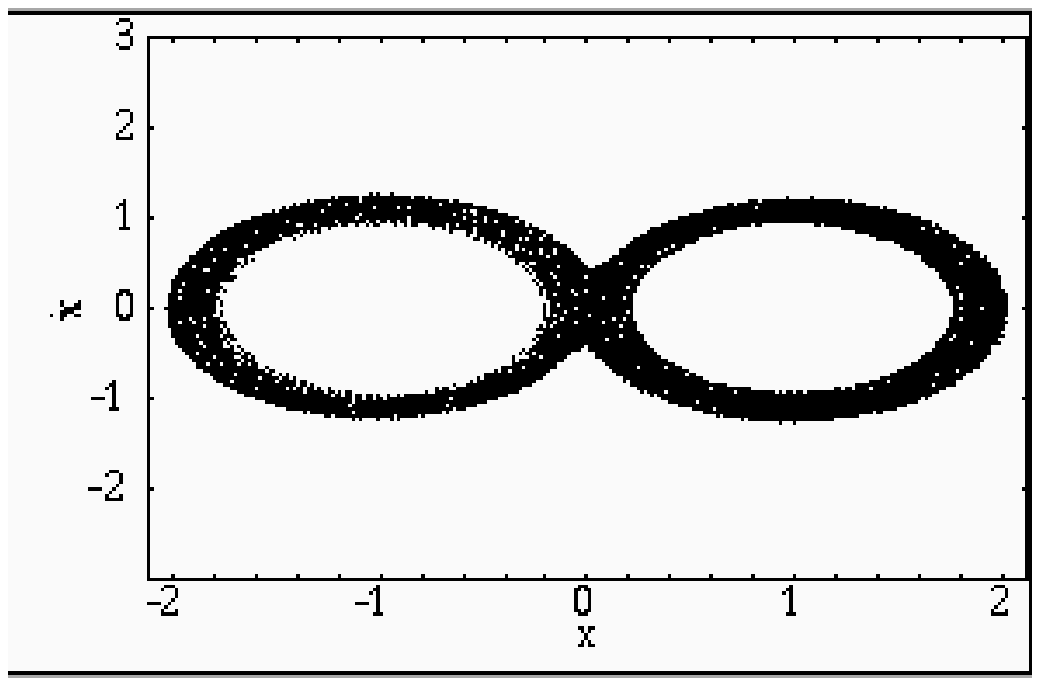,width=8.6cm} 
\\
FIG. \ref{xv-Lorenz-Lin}. R. Festa \emph{et al.}, Phys. Rev. E
\end{center}

\newpage
\begin{center} 
\psfig{file=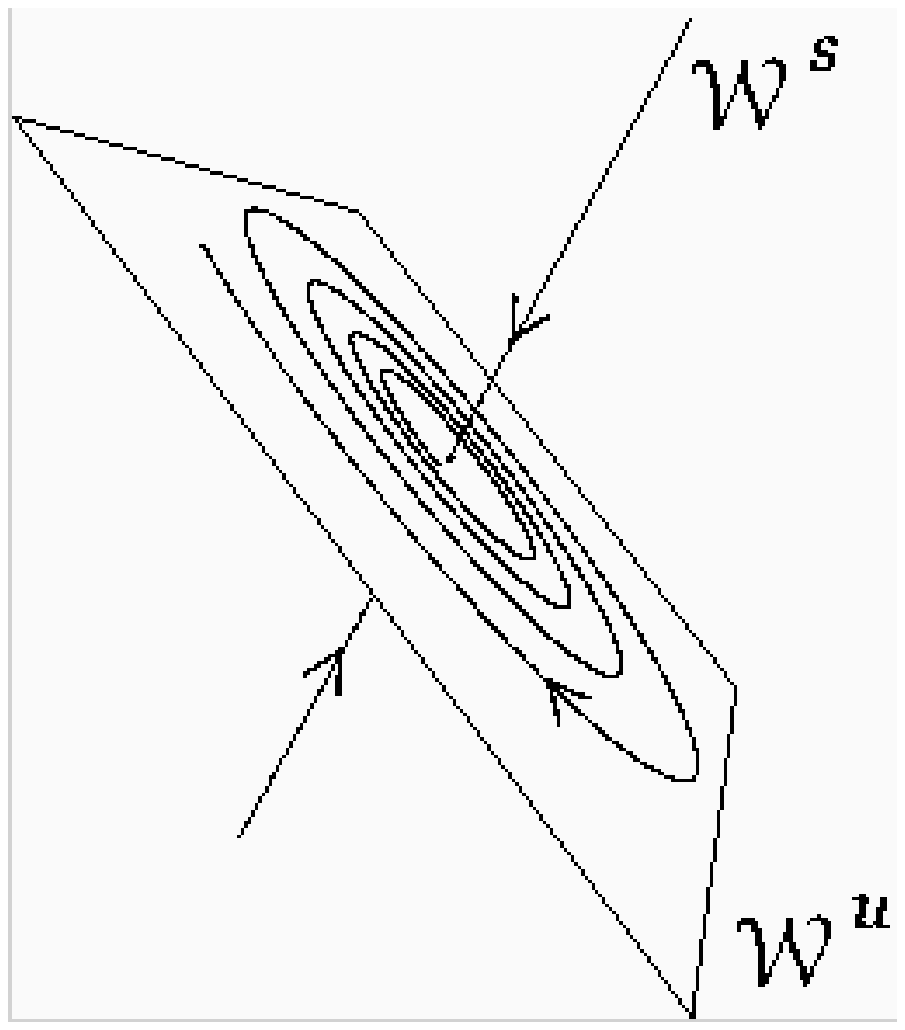,width=8.6cm} 
\\
FIG. \ref{sella}. R. Festa \emph{et al.}, Phys. Rev. E
\end{center}

\newpage
\begin{center} 
\psfig{file=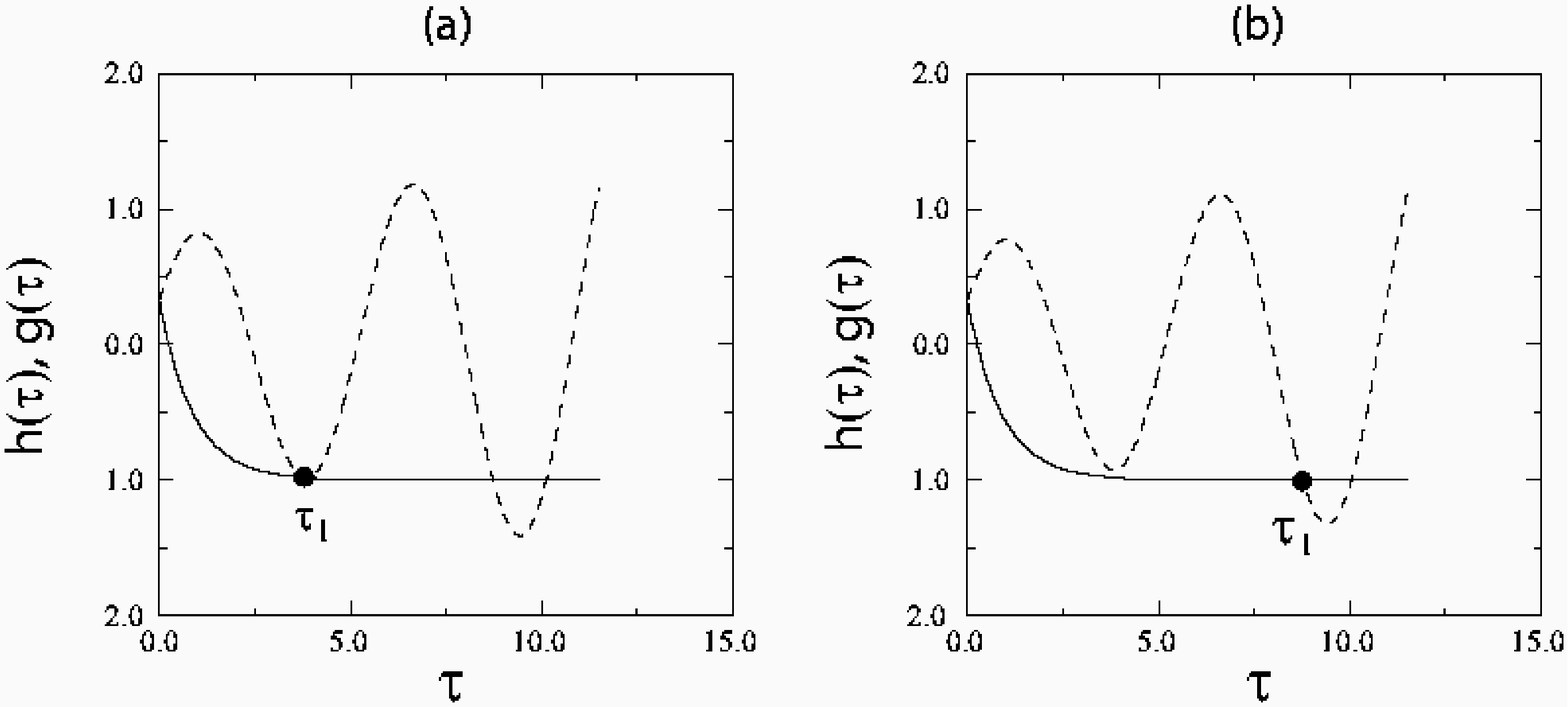,width=8.6cm} 
\\
FIG. \ref{intersez}. R. Festa \emph{et al.}, Phys. Rev. E
\end{center}

\newpage
\begin{center} 
\psfig{file=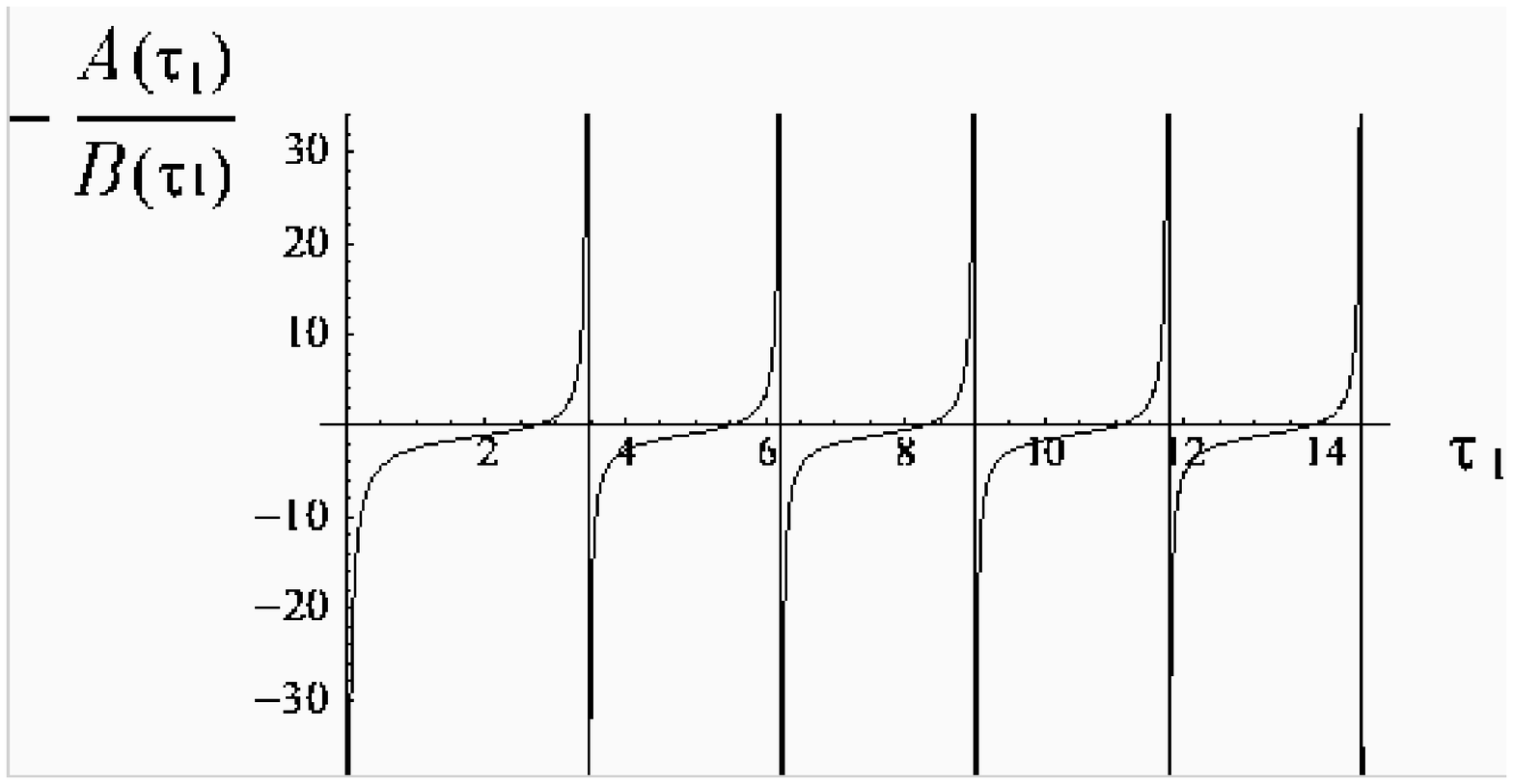,width=8.6cm} 
\\
FIG. \ref{coeff}. R. Festa \emph{et al.}, Phys. Rev. E
\end{center}

\newpage
\begin{center} 
\psfig{file=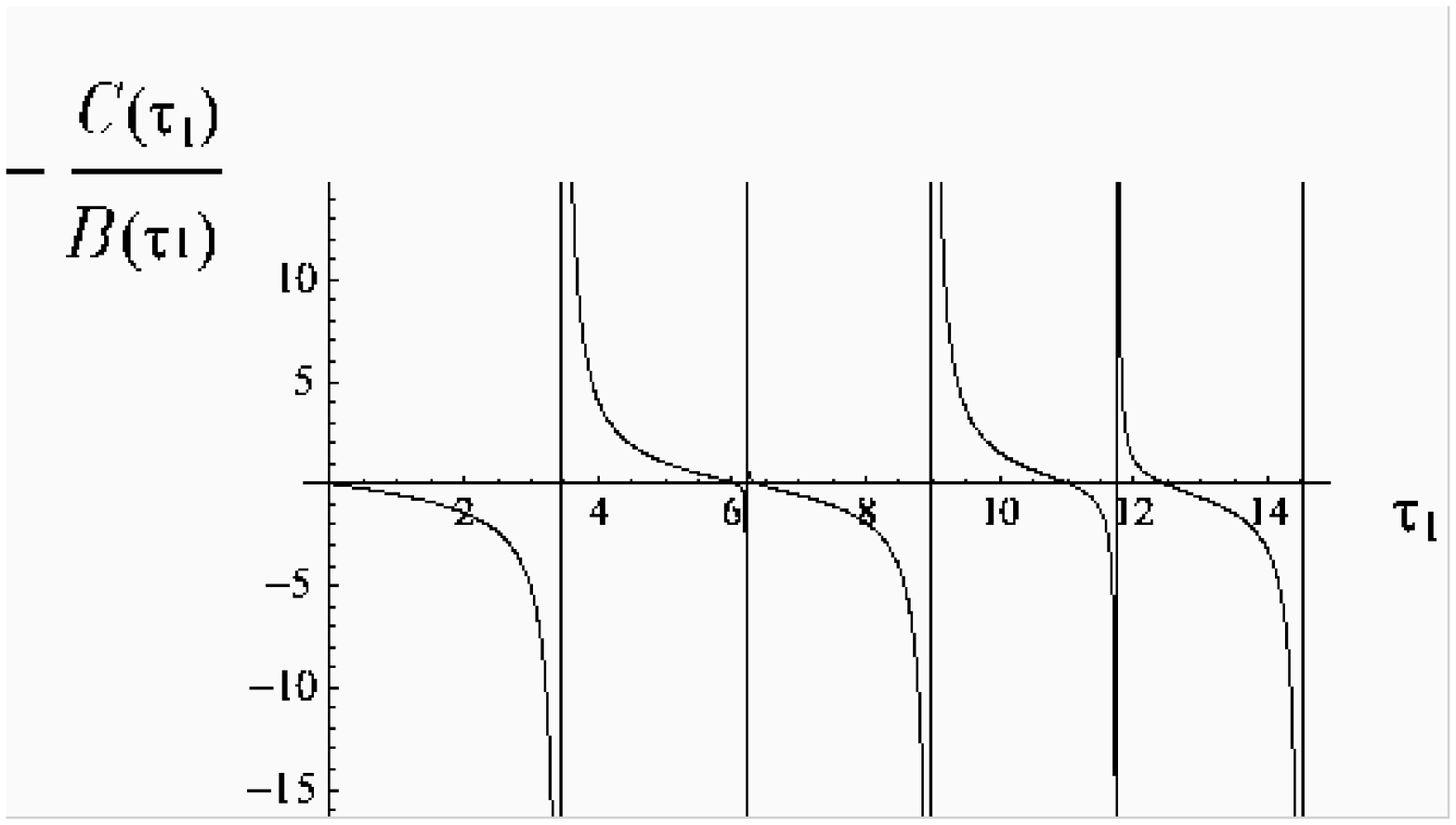,width=8.6cm} 
\\
FIG. \ref{intercetta}. R. Festa \emph{et al.}, Phys. Rev. E
\end{center}

\newpage
\begin{center} 
\psfig{file=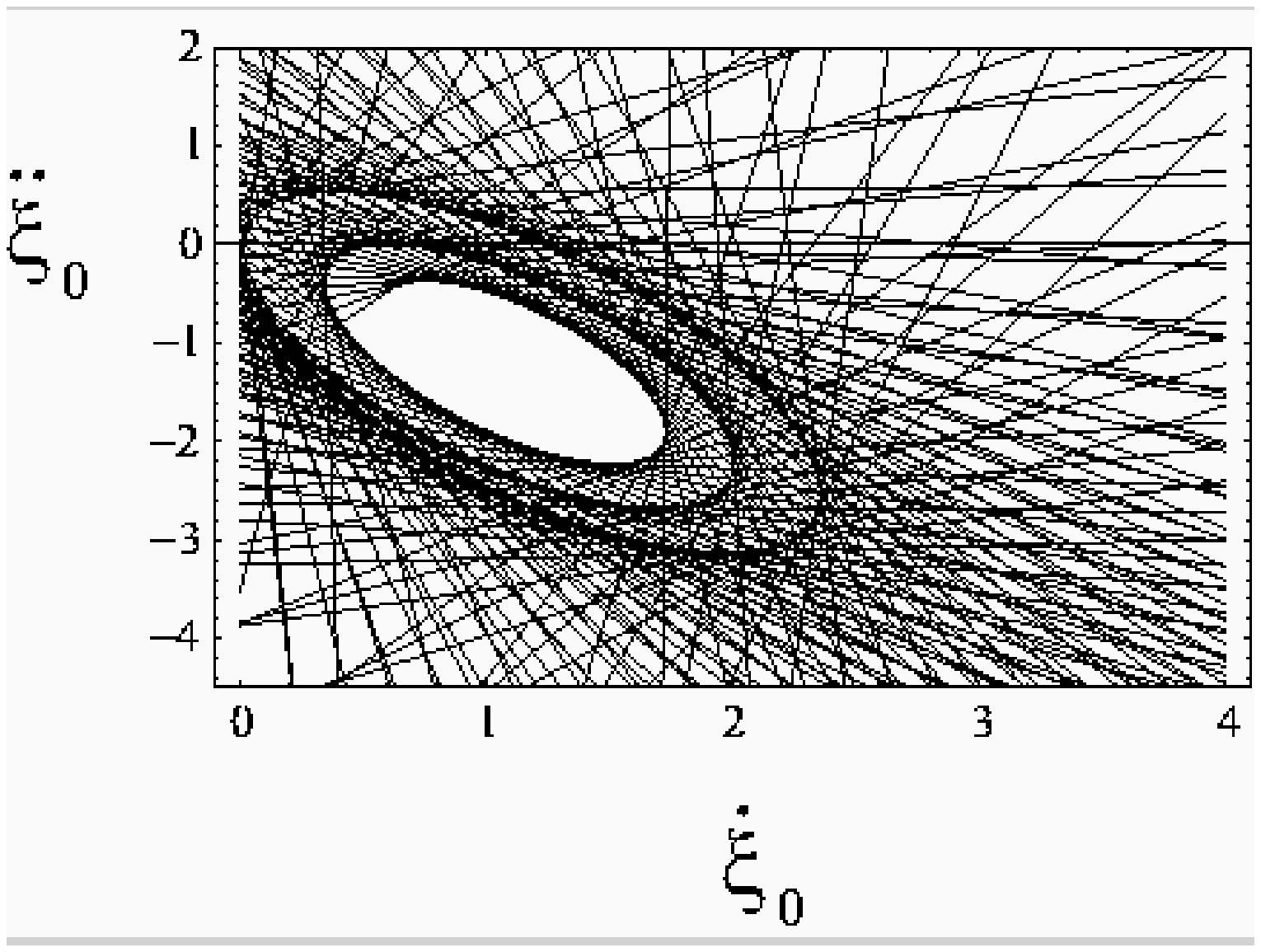,width=8.6cm} 
\\
FIG. \ref{ragnatela}. R. Festa \emph{et al.}, Phys. Rev. E
\end{center}

\newpage
\begin{center} 
\psfig{file=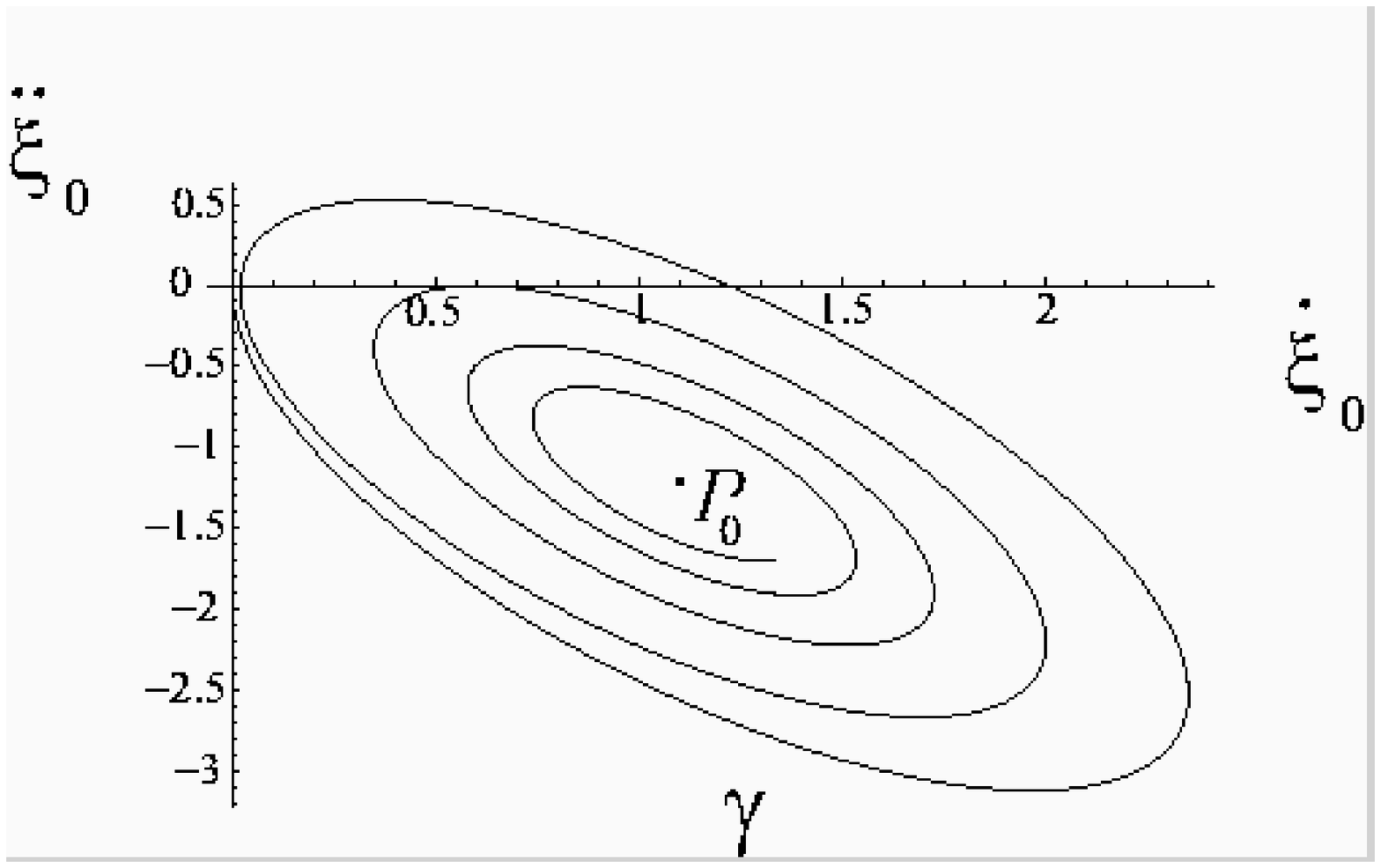,width=8.6cm} 
\\
FIG. \ref{inviluppo}. R. Festa \emph{et al.}, Phys. Rev. E
\end{center}

\newpage
\begin{center} 
\psfig{file=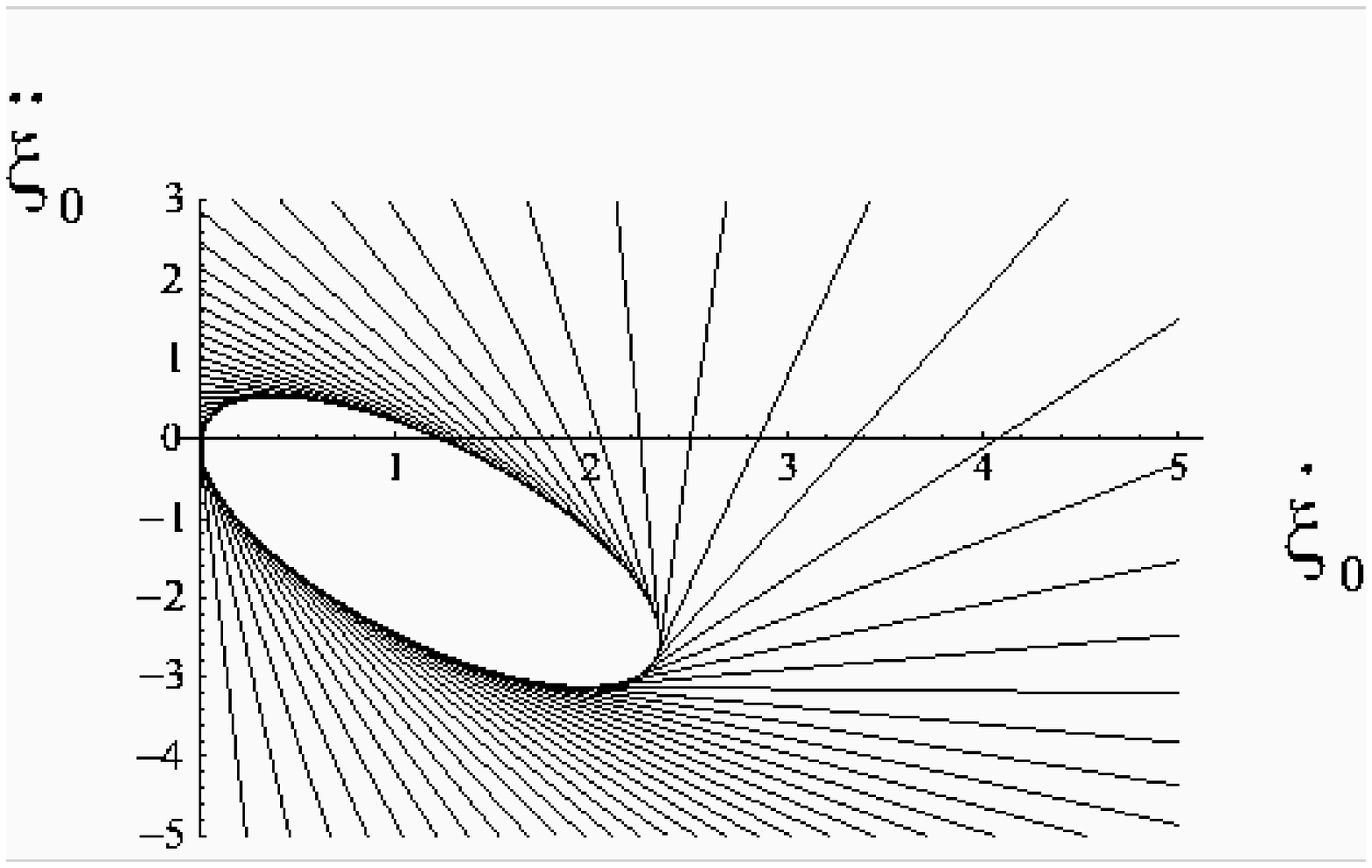,width=8.6cm} 
\\
FIG. \ref{semirette}. R. Festa \emph{et al.}, Phys. Rev. E
\end{center}

\newpage
\begin{center} 
\psfig{file=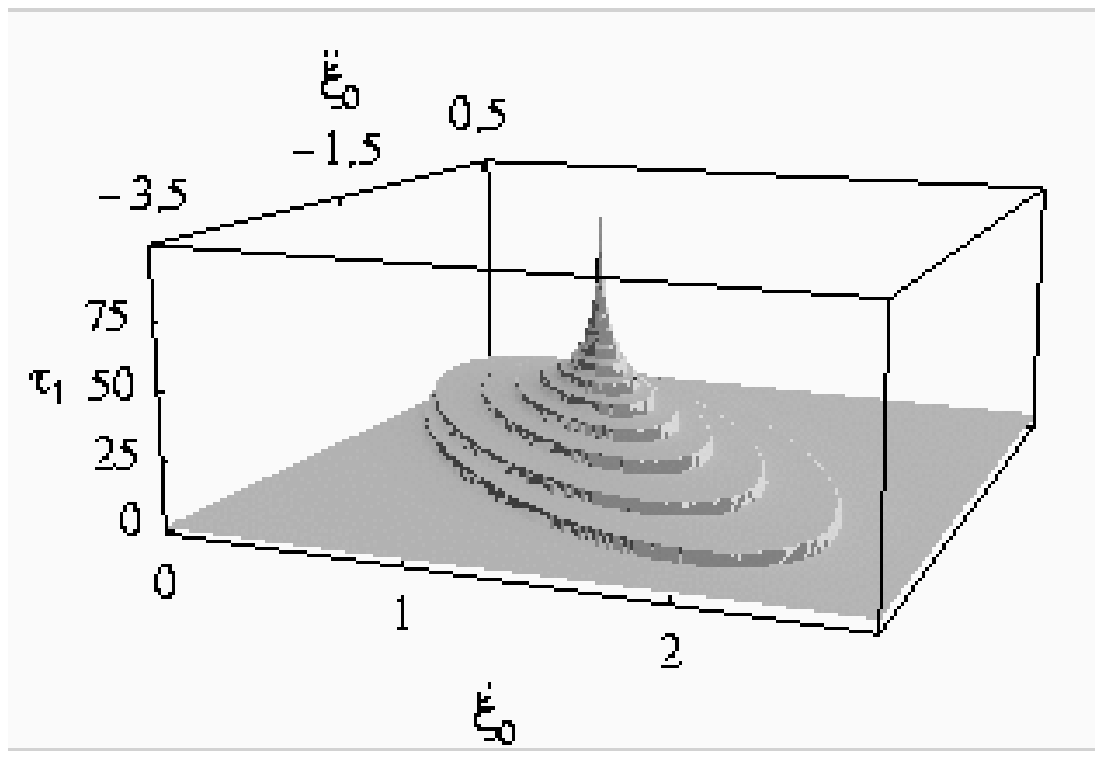,width=8.6cm} 
\\
FIG. \ref{piramide}. R. Festa \emph{et al.}, Phys. Rev. E
\end{center}

\newpage
\begin{center} 
\psfig{file=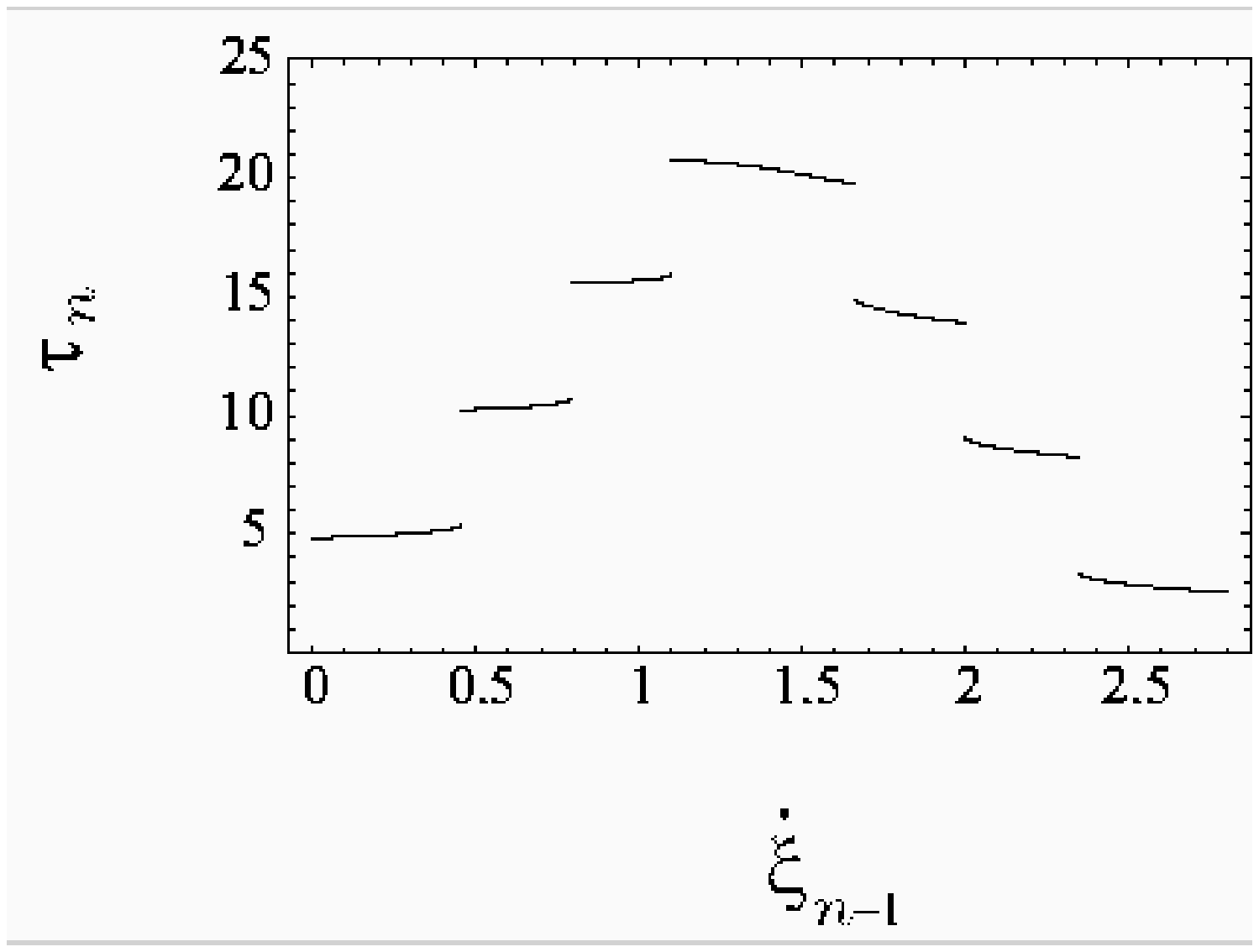,width=8.6cm} 
\\
FIG. \ref{t-v}. R. Festa \emph{et al.}, Phys. Rev. E
\end{center}

\newpage
\begin{center} 
\psfig{file=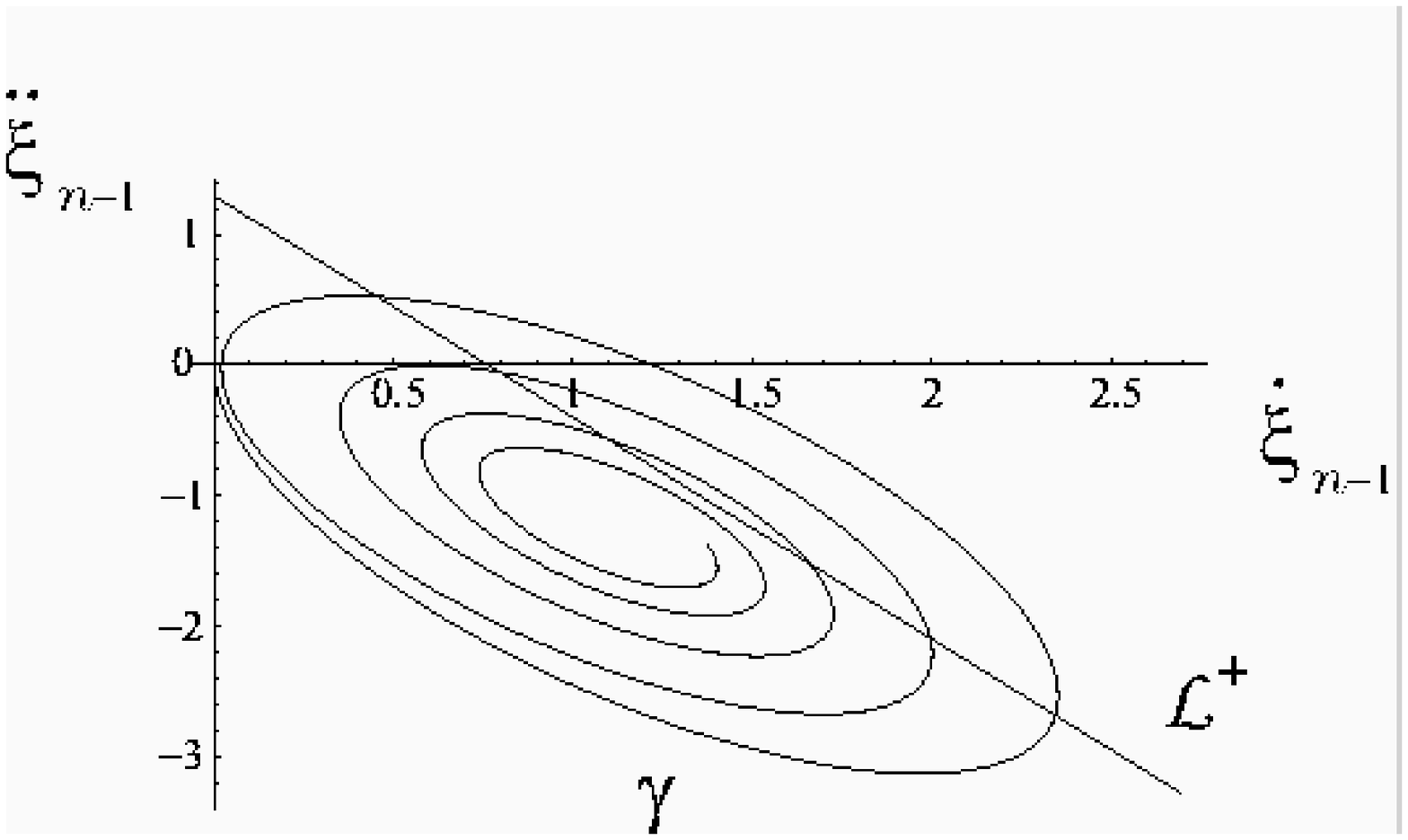,width=8.6cm} 
\\
FIG. \ref{sovrapp}. R. Festa \emph{et al.}, Phys. Rev. E
\end{center}

\newpage
\begin{center} 
\psfig{file=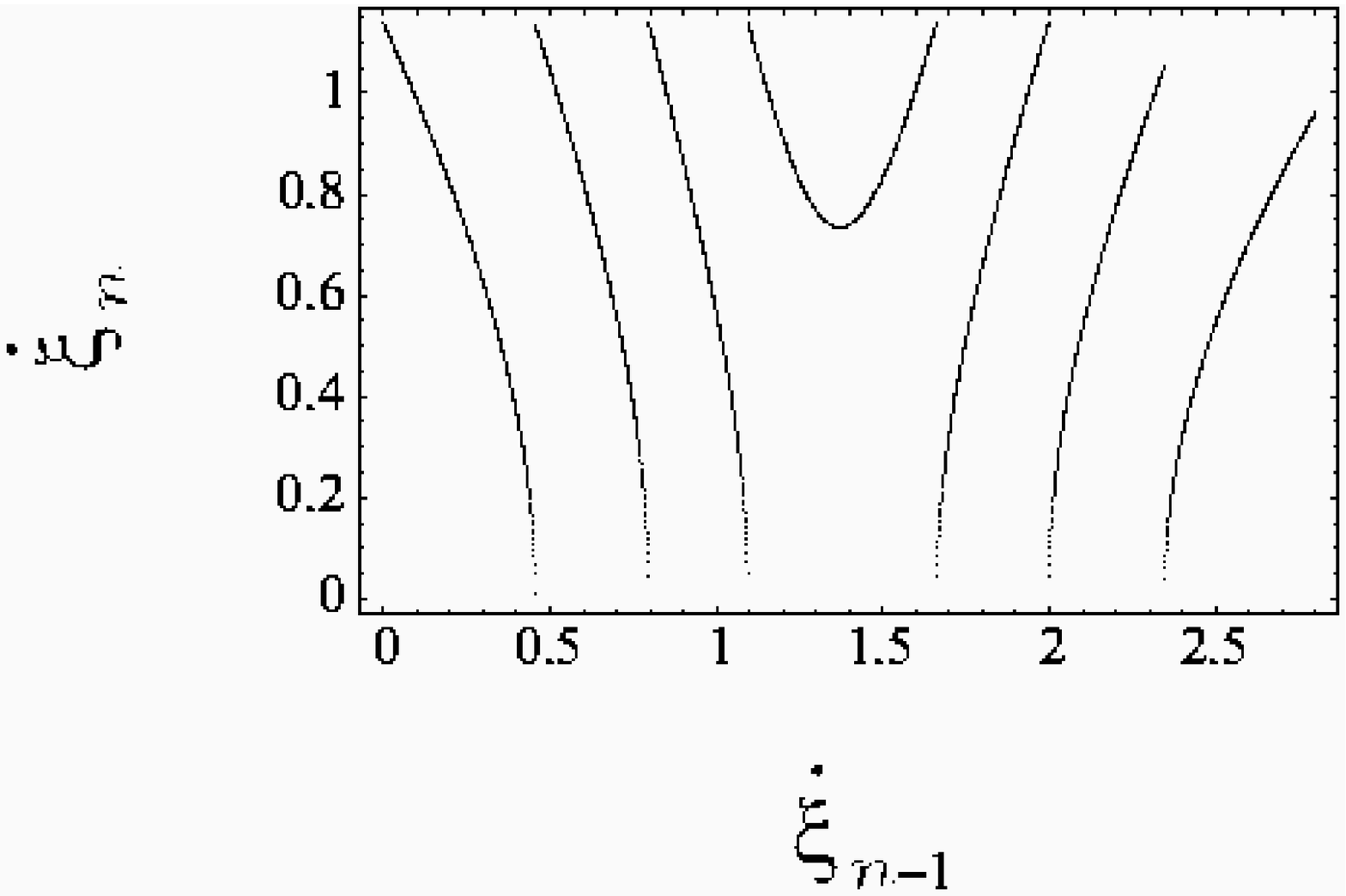,width=8.6cm} 
\\
FIG. \ref{v-v}. R. Festa \emph{et al.}, Phys. Rev. E
\end{center}

\newpage
\begin{center} 
\psfig{file=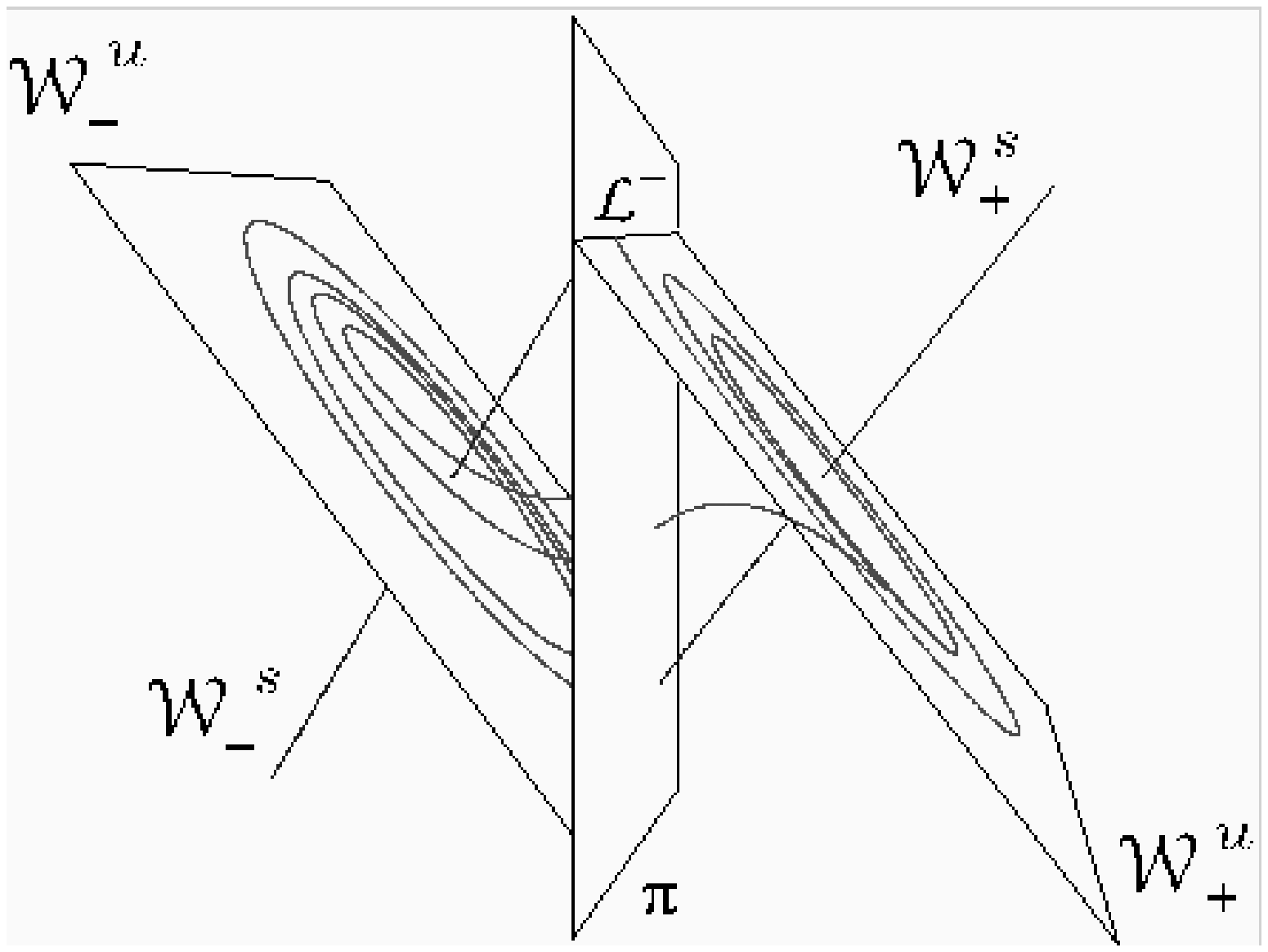,width=8.6cm} 
\\
FIG. \ref{spaziofasi}. R. Festa \emph{et al.}, Phys. Rev. E
\end{center}

\newpage
\begin{center} 
\psfig{file=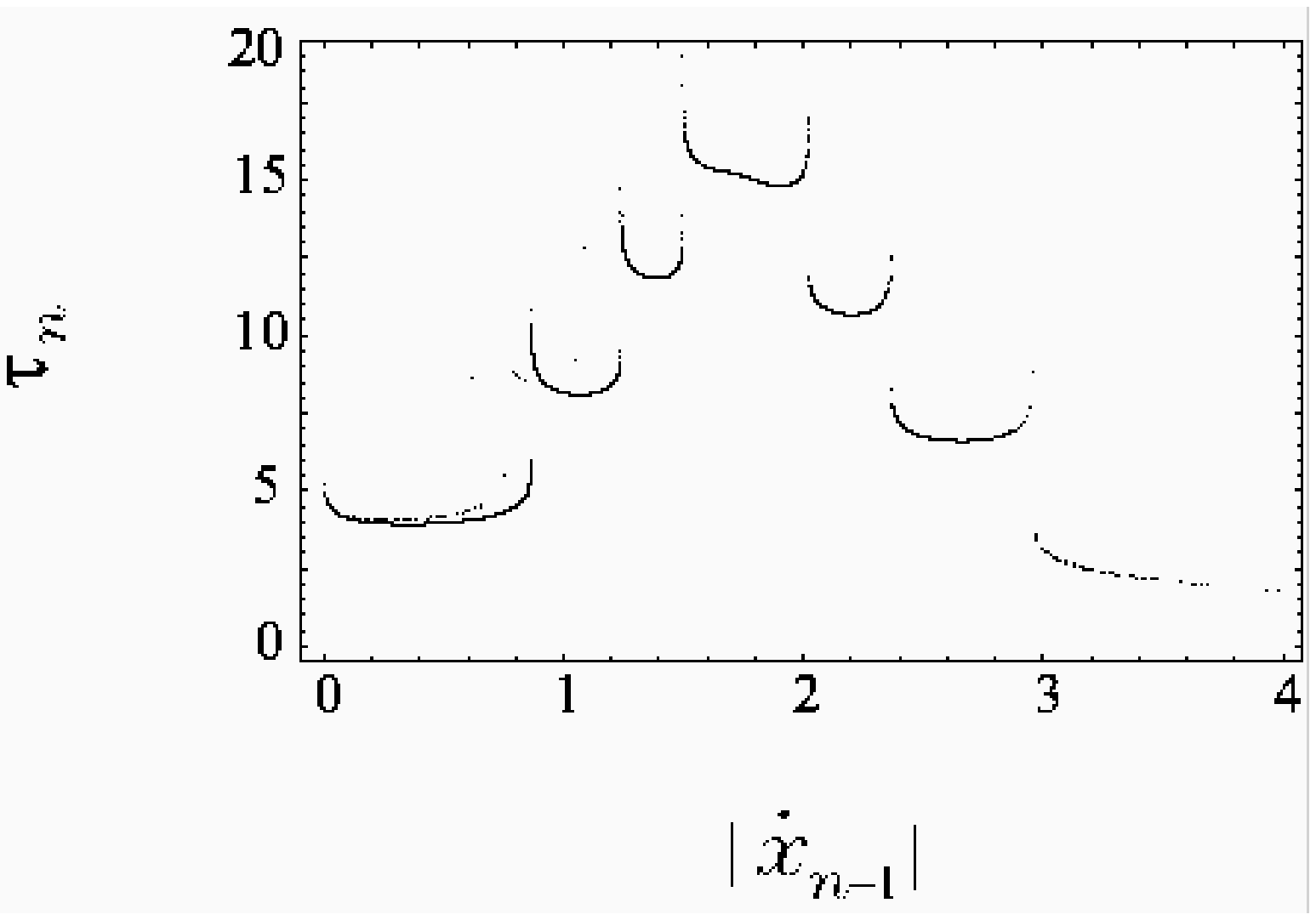,width=8.6cm} 
\\
FIG. \ref{t-vL}. R. Festa \emph{et al.}, Phys. Rev. E
\end{center}

\newpage
\begin{center} 
\psfig{file=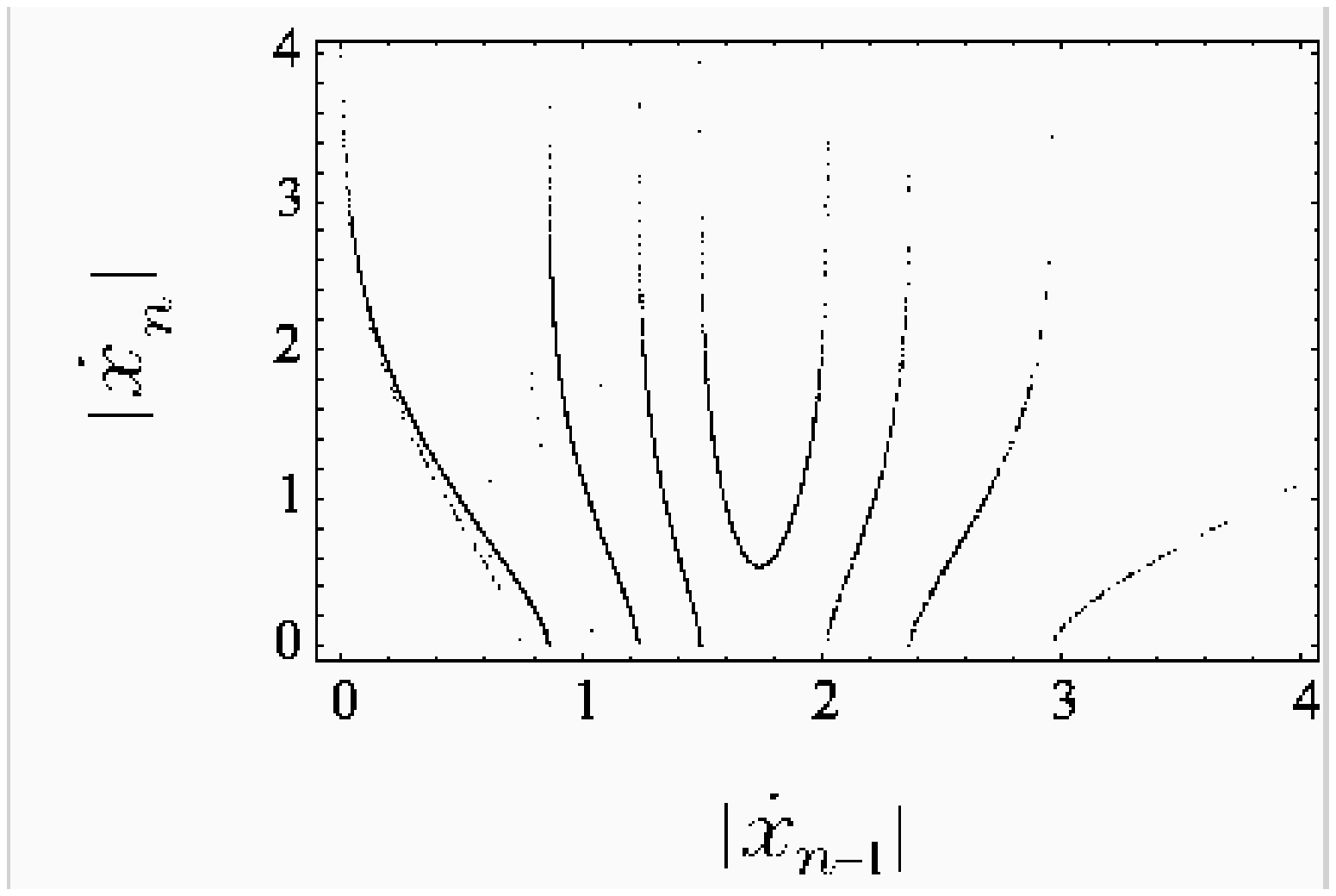,width=8.6cm} 
\\
FIG. \ref{v-vL}. R. Festa \emph{et al.}, Phys. Rev. E
\end{center}

\newpage
\begin{center} 
\psfig{file=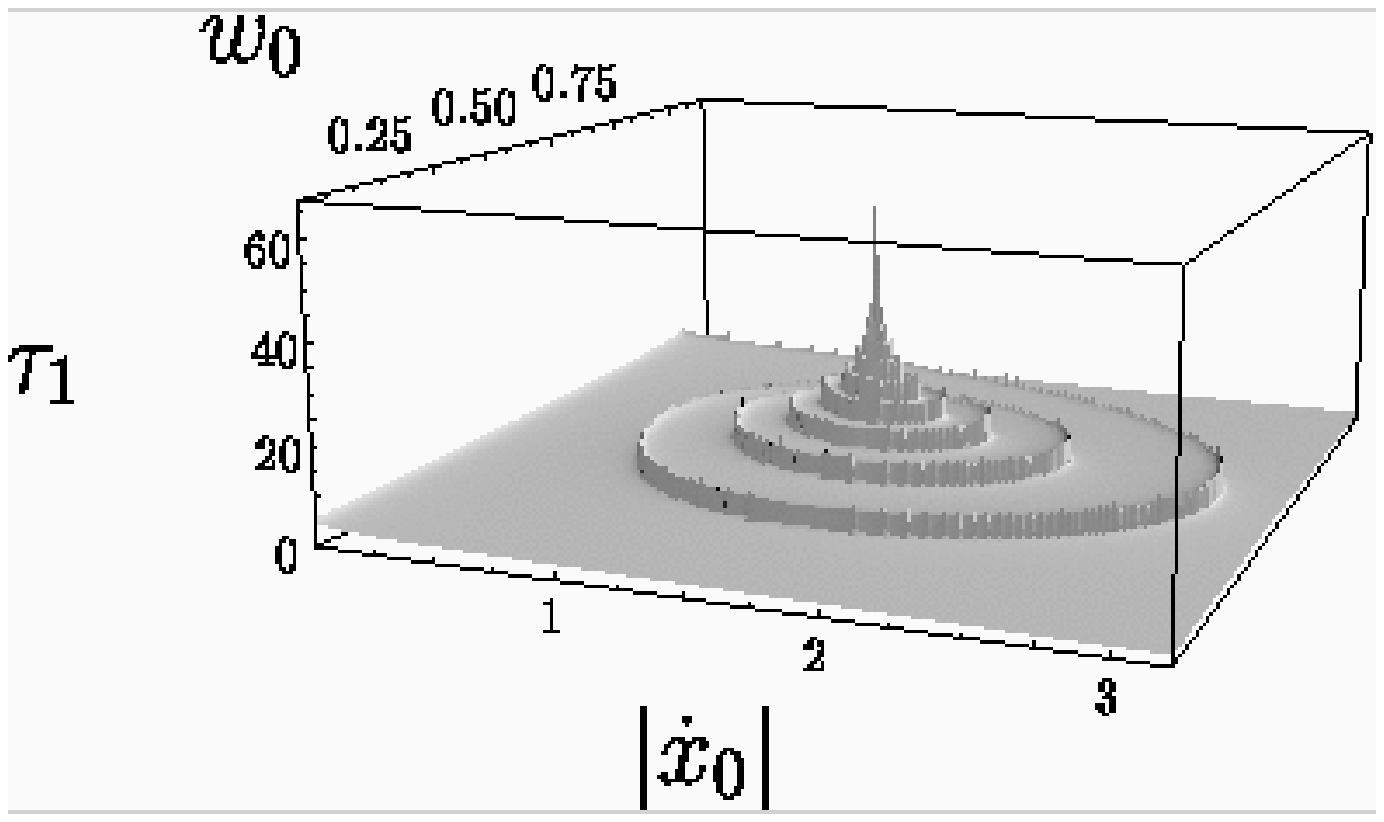,width=8.6cm} 
\\
FIG. \ref{torreL}. R. Festa \emph{et al.}, Phys. Rev. E
\end{center}

\newpage
\begin{center} 
\psfig{file=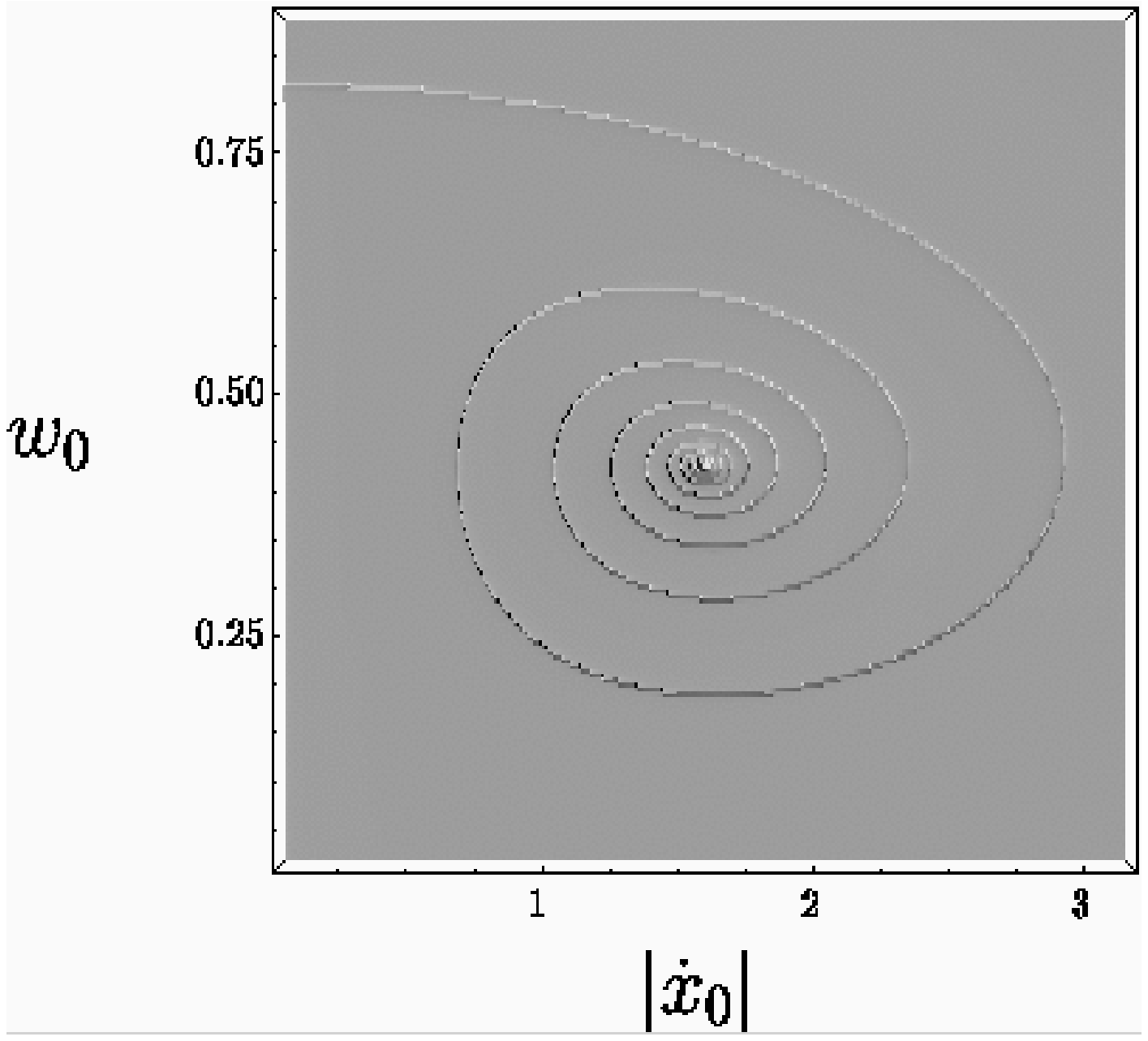,width=8.6cm} 
\\
FIG. \ref{torreL1}. R. Festa \emph{et al.}, Phys. Rev. E
\end{center}

\newpage
\begin{center} 
\psfig{file=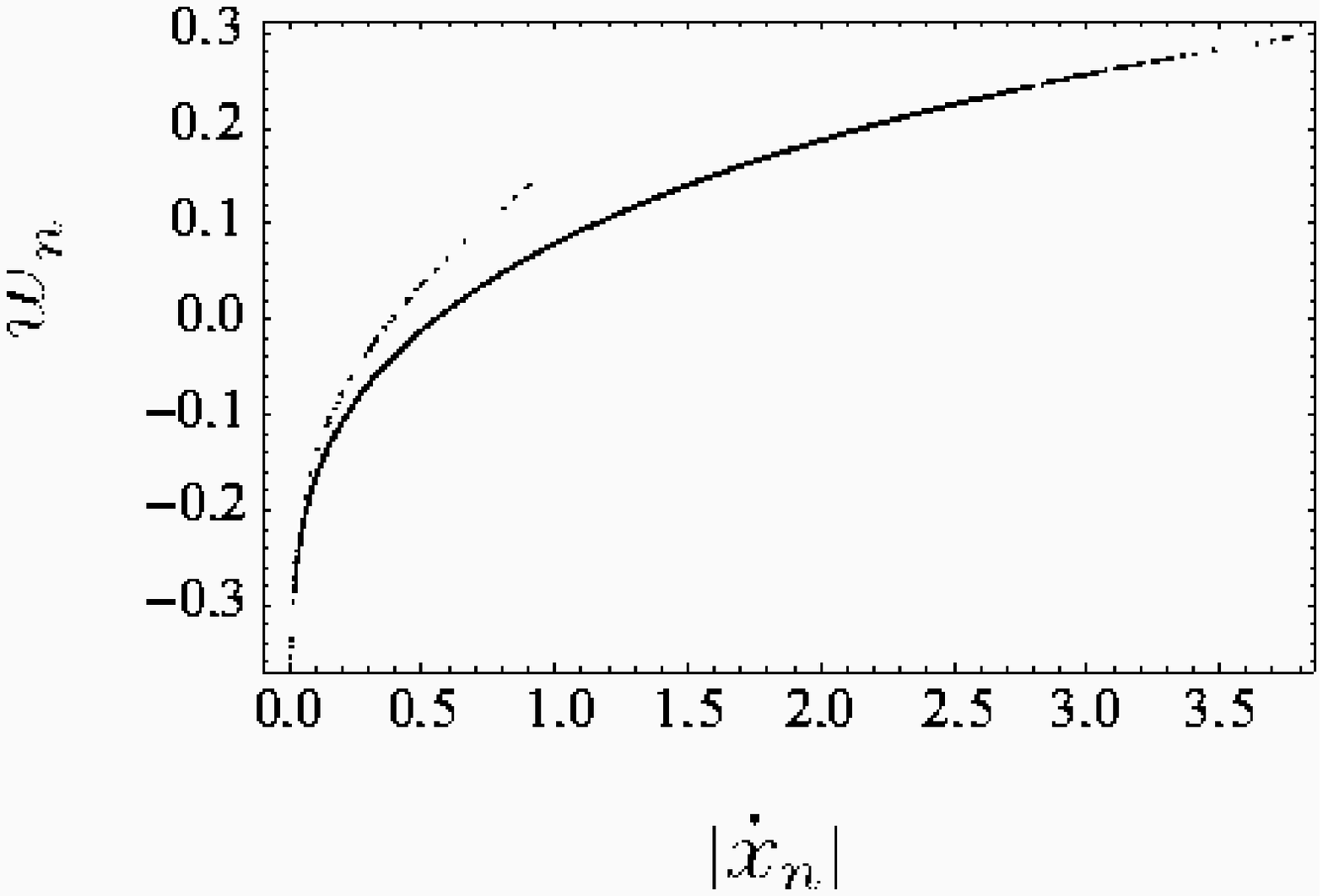,width=8.6cm} 
\\
FIG. \ref{memL}. R. Festa \emph{et al.}, Phys. Rev. E
\end{center}

\newpage
\begin{center} 
\psfig{file=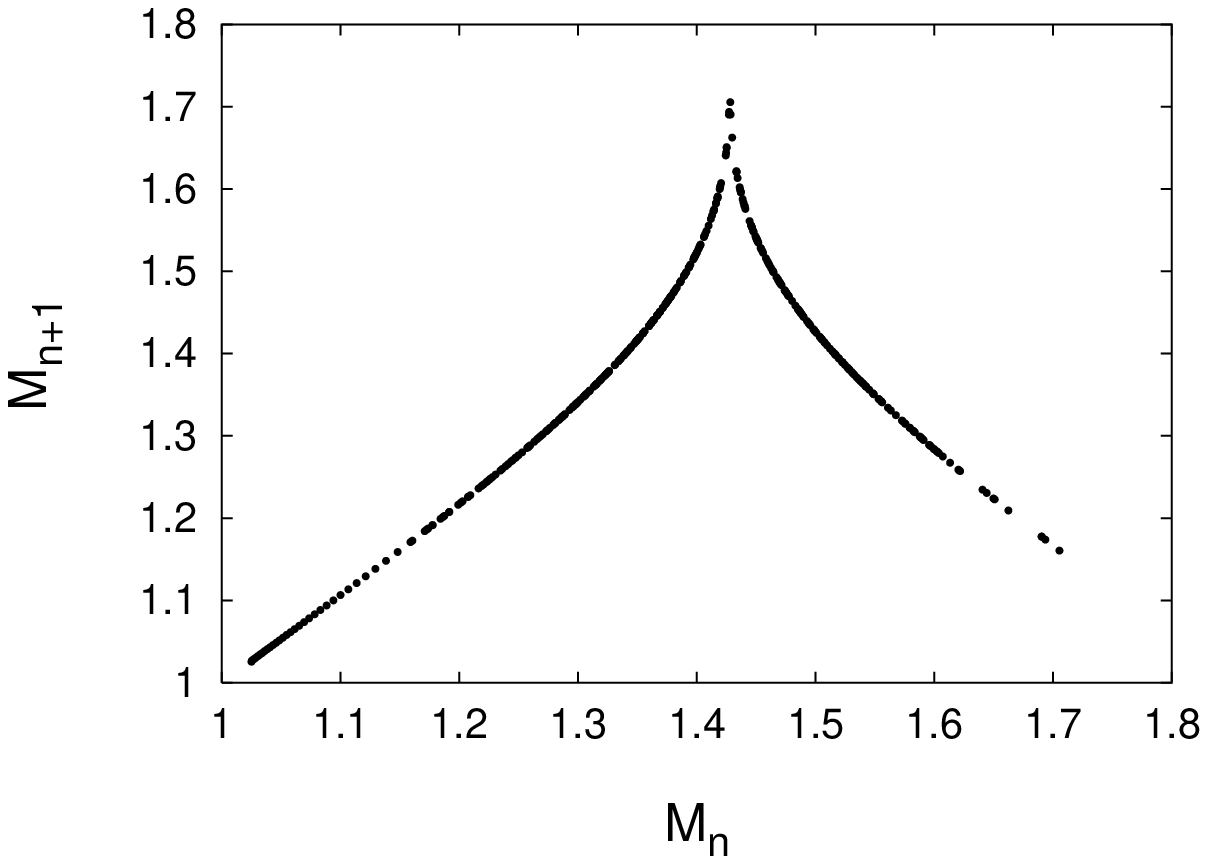,width=8.6cm} 
\\
FIG. \ref{tenda}. R. Festa \emph{et al.}, Phys. Rev. E
\end{center}
 
\newpage
\begin{center} 
\psfig{file=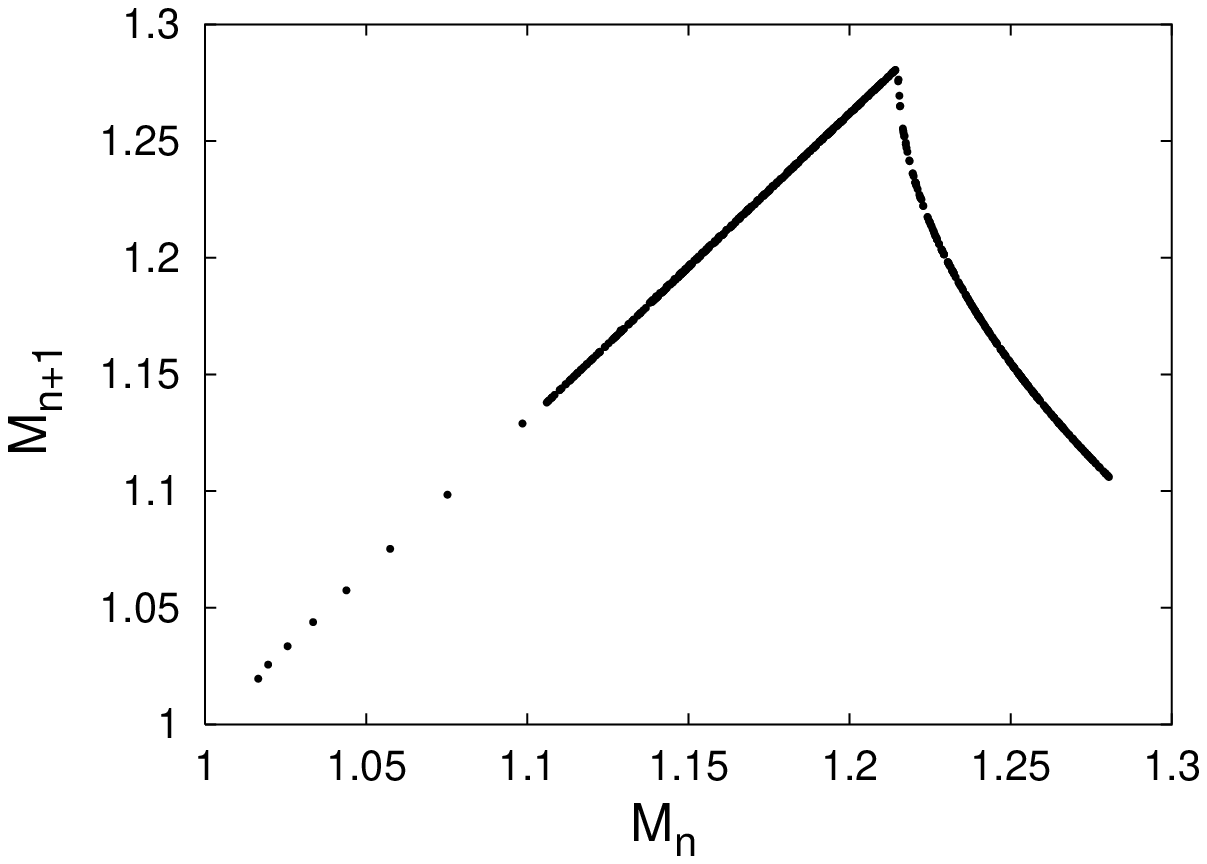,width=8.6cm} 
\\
FIG. \ref{tenda-lin}. R. Festa \emph{et al.}, Phys. Rev. E
\end{center}


\begin{thebibliography}{99}
\bibitem{nota:condiz.}
For the sake of simplicity we frequently speak of ``initial conditions''
for a trajectory after a generic collision. 
Obviously, we refer to the conditions that a solution, which
matches to a preceding trajectory, must satisfy in $x=0$.
We always consider the evolution of the system in the steady-state 
and so far from the initial transient.
\end{thebibliography}
\end{document}